\documentclass[aps,prb,twocolumn,superscriptaddress,10pt]{revtex4-2}
\usepackage{amsmath}
\usepackage{amsfonts}
\usepackage{amssymb}
\usepackage{array}% Extension to Table columns
\usepackage{bm}% bold math symbols
\usepackage{booktabs}% Three-line table
\usepackage{braket}% Dirac notation
\usepackage{cancel}% slash and backslash over symbols/texts
\usepackage{dcolumn}% align by a decimal point in table
\usepackage{graphicx}% Include figure files
\usepackage{hyperref}% Hyperlink support
\usepackage{mathtools}% Extension and additional support to AMS
\usepackage{multirow}% multirow and multicolumn support for Table
\usepackage{siunitx}% Typeset numbers and SI unit
\usepackage{spreadtab}% Draw table like spreadsheat
\usepackage[caption=false]{subfig}% subfig plotting, Fig.1(a), (b), etc
\usepackage{xcolor}% Text color support
\usepackage{tabularx} % Include table
\usepackage[noabbrev]{cleveref}
\usepackage{threeparttablex}

\usepackage{pax} % add links  of the included pdfs
\hypersetup{
breaklinks=true,   % splits links across lines
colorlinks=true,    % displays links as colored text instead of blocks
linkcolor=blue,
citecolor=blue,
urlcolor=blue,
}
\graphicspath{{./figures/}}		%figures are located in the path
\newcolumntype{d}[1]{D{.}{.}{#1}}

\usepackage{pdfpages} % include pdfs
\usepackage{pgffor} % for loops

\makeatletter
\AtBeginDocument{\let\LS@rot\@undefined}
\makeatother

\def\supplementfilename{SI.pdf}

\pdfximage{\supplementfilename}
\def\numbersupplementpages{\the\pdflastximagepages}

\newif\ifarXiv
\arXivtrue 
\begin{document}

%\title{Phonon physics from first principles using the \texttt{Pheasy} code}
\title{First-principles phonon physics using the \texttt{Pheasy} code}

\author{Changpeng Lin}
\email{changpeng.lin@epfl.ch}
\affiliation{%
Theory and Simulation of Materials (THEOS), and National Centre for Computational Design and Discovery of Novel Materials (MARVEL), \'Ecole Polytechnique F\'ed\'erale de Lausanne, 1015 Lausanne, Switzerland
}%
\author{Jian Han}
\affiliation{%
State Key Laboratory of New Ceramics and Fine Processing, School of Materials Science and Engineering, Tsinghua University, 100084 Beijing, People's Republic of China
}%
\author{Ben Xu}
\email{bxu@gscaep.ac.cn}
\affiliation{Graduate School of China Academy of Engineering Physics, 100088 Beijing, People's Republic of China}%
\author{Nicola Marzari}
\affiliation{%
Theory and Simulation of Materials (THEOS), and National Centre for Computational Design and Discovery of Novel Materials (MARVEL), \'Ecole Polytechnique F\'ed\'erale de Lausanne, 1015 Lausanne, Switzerland
}%
\affiliation{%
Laboratory for Materials Simulations, Paul Scherrer Institut, 5232
Villigen PSI, Switzerland
}%

\date{\today}

\begin{abstract}
Parameter-free calculations of lattice dynamics from first principles have achieved significant progress in the past decades, with a wealth of applications in thermodynamics, phase transitions, and transport properties of materials.
Current approaches to derive the interatomic force constants (IFCs) of lattice potential become challenging and sometimes infeasible when going beyond third-order anharmonicity, due to the combinatorial explosion in the number of higher-order IFCs.
In this work, we present a robust and user-friendly program, \texttt{Pheasy}, which accurately reconstructs the potential energy surface of crystalline solids via a Taylor expansion of arbitrarily high order.
Given force-displacement datasets, the program enables an efficient and accurate extraction of IFCs using advanced machine-learning algorithms, and further calculates a wide range of harmonic and anharmonic phonon related properties.
We show in three prototypical examples how the obtained IFCs have been successfully applied to study anharmonic lattice dynamics and thermal transport.
Through these detailed benchmarks, we have also identified the optimal approach for IFC extractions and offered general guidelines for high-fidelity lattice-dynamical simulations, addressing the large uncertainties in the IFCs extracted from existing various schemes.
Overall, the \texttt{Pheasy} project aims to create a phonon code ecosystem that connects diverse phonon simulation platforms and offers access to the broad research community.
\end{abstract}

\maketitle

\section*{Introduction}
The study of atomic vibrations in solids has been one of the cornerstones in modern condensed matter physics and allows to understand or predict manifold physical phenomena that underpin technological advancements, including superconductivity~\cite{Grimvall1981,Giustino2017}, optical processes~\cite{Ridley2013,Chow1969,Noffsinger2012}, phase transitions~\cite{Dove1993,Cowley1964}, transport properties~\cite{Ziman1960,Ponce2020,Claes2025,Qian2021} and many others.
For example, the sound propagation in a medium is a direct manifestation of collective acoustic vibrations, and the strong absorption of a certain infrared light confirms the existence of specific atomic motions actively coupled to electromagnetic fields.
Early studies on lattice dynamics stemmed from the attempts to understand the observed deviation from the Dulong--Petit law~\cite{Dulong1819} in the low-temperature heat capacity of solids, where the first quantum theory was formulated by Einstein~\cite{Einstein1907} in 1907 and further extended by Debye~\cite{Debye1912}.
The formal microscopic treatment on realistic three-dimensional crystals is owing to Born and Von K{\'a}rm{\'a}n~\cite{Born1912,Born1913}, which lays the foundation to the modern theory of lattice dynamics.
They employed a model crystal whose atoms interact with the surrounding neighbors via spring-like forces, known as the harmonic approximation, and the well-known Born--Oppenheimer approximation~\cite{Born1927} was also assumed to decouple the electronic and ionic degrees of freedom in solids.
Within the Born--Von K{\'a}rm{\'a}n theory~\cite{Born1912,Born1913,Born1954}, the quantization of their atomic eigenmotions yields phonons that are quasiparticles with a well-defined dispersion relation.
Notably, in such harmonic approximation, phonons are non-interacting, possessing a definite energy but an infinite lifetime.
The harmonic theory of lattice dynamics has been proved quite successful in explaining many physical properties of solids, such as thermodynamic functions~\cite{Mounet2005,Fultz2010,Togo2010} and phonon-mediated carrier dynamics~\cite{Giustino2017,Noffsinger2012,Ponce2020,Claes2025}.
However, there are also many exceptions where anharmonic effects are essential to correctly describe their physical behaviors.

Lattice anharmonicity plays a crucial role in interpreting plenty of phenomena where the standard harmonic approximation fails, including thermal expansion~\cite{Maradudin1962b,Togo2010,Ritz2019}, thermal transport~\cite{Qian2021,Broido2007,Li2014} and structural phase transitions~\cite{Dove1993,Cowley1964}.
Consequently, phonons are no longer independent, and prevalent phonon-phonon interactions give rise to their frequency shifts and finite lifetimes; these correspond to the physical observables that can be directly measured in infrared, Raman and inelastic neutron/X-ray scattering spectroscopies~\cite{Maradudin1962,Squires2012}.
When anharmonic effects are small compared to harmonic contributions (i.e. weakly anharmonic systems), thermal expansion effect is often tackled within the so-called quasi-harmonic approximation~\cite{Togo2010,Ritz2019}, which only takes into account the volume dependence of phonon frequencies.
Similarly, phonon-phonon collisions are built on the top of harmonic phonons, and only the leading three-phonon (3ph) scattering is calculated using the first-order perturbation theory~\cite{Broido2007,Li2014,Ziman1960}.
It is important to note that such perturbative treatments have neglected the explicit temperature dependence in phonon frequencies.
Indeed, perturbative approaches collapse for strongly anharmonic systems, which show a significant temperature-dependent phonon dispersion and, in many cases, even exhibit a structural instability within the harmonic approximation.
In order to overcome the limitations of harmonic approximation and perturbation theory, many non-perturbative approaches have been proposed, such as self-consistent harmonic approximation (SCHA)~\cite{Hooton1955a,Souvatzis2008,Monacelli2021}, self-consistent phonons (SCP)~\cite{Gillis1968,Goldman1968,Werthamer1970,Tadano2015} and temperature-dependent effective potentials (TDEP)~\cite{Hellman2011,Hellman2013a,Hellman2013b}.

The central quantities in lattice-dynamical theory are interatomic force constants (IFCs), which are the derivatives of Born--Oppenheimer potential energy surface with respect to atomic displacements.
Particularly, the second-order IFCs define phonon dispersion relations, whereas the third- and high-order IFCs govern anharmonic phonon-phonon interactions.
A major breakthrough was the development of density functional perturbation theory (DFPT)~\cite{Baroni1987,Giannozzi1991,Baroni2001,Gonze1997a} and ever since, advances in density functional theory (DFT) have enabled parameter-free calculations of phonon dispersions for an arbitrary material, available in many open-source software packages.
There are essentially two approaches for evaluating the second-order IFCs from first principles: the real-space small displacement (or frozen-phonon) method~\cite{Kunc1982,Parlinski1997,Alf2009} and the aforementioned reciprocal-space linear response method within DFPT~\cite{Baroni1987,Giannozzi1991,Baroni2001,Gonze1997a}; the former is implemented, e.g., in the \texttt{Phon}~\cite{Alf2009} and \texttt{Phonopy}~\cite{Togo2015,Togo2023a,Togo2023b} codes, while the latter is available in the \texttt{Quantum ESPRESSO}~\cite{Giannozzi2009,Giannozzi2017} and \texttt{Abinit}~\cite{Gonze2016,Gonze2020} packages.
Nonetheless, the calculations of anharmonic IFCs and beyond using these conventional approaches are challenging and sometimes become unaffordable, due to the combinatiorial explosion in the number of high-order IFCs, especially when going beyond the third-order anharmonicity.
The generalization of small displacement method to high-order derivatives is indeed straightforward and systematic, where multiple atom sites are perturbed simultaneously for a single displaced configuration. 
Such implementations are mainly limited to lowest third-order anharmonicity for thermal transport simulations, as exemplified in \texttt{thirdorder.py}~\cite{Li2014} and \texttt{Phono3py}~\cite{Togo2023a,Togo2023b}.
Although its further extension to the next fourth-order IFCs become available in the recent \texttt{fourthorder.py} code~\cite{Han2022}, significant computational costs due to the exponential surge in the number of required displaced configurations have restricted practical applications to high-symmetry simple crystals with a small IFC cutoff radius.
Likewise, linear response approach encounters the same numerical difficulties, whose implementation is further complicated by its pseudopotential dependence and necessities a specialized code~\cite{Baroni2001,Gonze1997a,DalCorso1997}; the current implementation supports only third-order derivatives in the \texttt{D3Q} code~\cite{Paulatto2013}.
Therefore, a systematic and computationally efficient approach to extract the high-order expansion coefficients of lattice potential from first principles is highly desirable.

Thanks to the introduction of linear-regression-based supercell approaches, the past decade has witnessed several significant progress in the calculation of high-order derivatives beyond cubic anharmonicity.
One prominent development is the compressive-sensing lattice dynamics (CSLD)~\cite{Zhou2014,Zhou2019a,Zhou2019b} which leverages a technique, originally used for reconstructing a signal from underdetermined linear systems~\cite{Candes2008}, to efficient build a sparse representation of lattice-dynamical models.
With a carefully selected regularization parameter for the additional $\ell_1$ penalty term, CSLD guarantees the sparsity of the high-order IFCs and avoids the common overfitting issues of linear regression.
The success of the CSLD technique has the root in the physical reality that IFCs of any system are generally sparse and decay rapidly with increasing interatomic distance, in particular for high-order anharmonic terms.
One caveat should be noted for infrared-active solids, where the interatomic interactions are inherently long-ranged owing to the vibration-induced macroscopic electric fields.
Since electrostatic contributions are defined analytically in the long-wavelength limit, a standard remedy based on the Ewald summation technique~\cite{Baroni2001,Gonze1997a} can be applied to separate the long-range and short-range components of interatomic forces~\cite{Zhou2019b}; the lowest-order approximation is to remove only dipole-dipole interactions, and the resulting short-range IFCs are ensured to decay faster than $1/d^3$~\cite{Gonze1997a} ($d$ being the interatomic distance).
CSLD has been implemented in several phonon packages, including the CSLD code from its original authors~\cite{Zhou2014,Zhou2019a,Zhou2019b}, \texttt{Alamode}~\cite{Tadano2014,Tadano2015} and \texttt{hiPhive}~\cite{Eriksson2019}.

In this work, we present the \texttt{Pheasy} code, a \texttt{Python} package for robust and efficient phonon calculations from first principles.
While this code can efficiently extract high-order IFCs of Born-Oppenheimer potential using machine-learning algorithms, it also provides a broad spectrum of phonon properties, enabling the studies of phonon-mediated phenomena.
For instance, with second-order IFCs, the code calculates phonon spectra and most of thermodynamic properties.
Notably, we incorporate the complete set of invariance conditions on harmonic IFCs and a dimension-dependent treatment for long-range Coulomb interactions, which are important for recovering the physical quadratic dispersions of flexural acoustic (ZA) modes~\cite{Lin2022} and the correct LO-TO splitting in low dimensions~\cite{Sohier2017,Royo2021}.
Besides, thanks to our recent developments~\cite{Lin2024}, long-range electrostatic interactions in the phonon dispersions of semiconductors are now addressed up to the second order of phonon momentum in the long-wavelength limit, beyond the lowest-order dipole-dipole approximation.
Specifically, the code further includes dynamical quadrupolar and octupolar effects that are crucial for accurately interpolating and converging the phonon dispersions of piezoelectric materials.
These novel developments make our code stand out from existing supercell-based phonon software, complementing the phonon community.
Furthermore, using anharmonic IFCs, the code can perform finite-temperature structure optimization considering both thermal expansion and thermal fluctuations of ionic positions, as well as the calculations of anharmonic renormalized phonon spectra using the SCHA and SCP techniques.
The detailed description and results of anharmonic lattice dynamics will be presented in a separate paper.

The aim of the \texttt{Pheasy} project is to build a user-friendly and ecosystem phonon code, which seamlessly integrates with various phonon simulation platforms.
The current code interfaces with \texttt{ShengBTE}~\cite{Li2014,Han2022}, \texttt{Phono3py}~\cite{Togo2023a,Togo2023b}, \texttt{Phoebe}~\cite{Cepellotti2022} and \texttt{GPU\_PBTE}~\cite{Zhang2021} for lattice thermal conductivity (LTC) calculations, and with \texttt{EPW}~\cite{Lee2023} for phonon-mediated carrier mobility calculations with anharmonic effects included.
Given the variety of linear-regression-based IFC extraction flavors in the literature, it is imperative to identify the optimal scheme for reliable lattice-dynamical and thermal transport calculations.
Thus, we here rigorously demonstrate the most appropriate scheme for extracting high-order IFCs, through a careful benchmark of the calculated anharmonic phonon dispersions.
We also provide practical guidelines for IFC extractions and highlight potential pitfalls, which ensure accurate and reliable phonon simulations in the future.
Last but not least, besides the benchmark calculations presented in this work, the \texttt{Pheasy} code has been extensively utilized in many studies~\cite{Lin2022,Feng2022,Lin2022,Rodriguez2023a,Rodriguez2023b,Zheng2024,AlFahdi2025,Wei2025,Xiong2024,Han2023,Zheng2023,Ganose2025}, further validating its correctness and soundness.

\section*{Results}

In this section, we first outlines the general theory of lattice dynamics and its implementation in the \texttt{Pheasy} code.
The accuracy and reliabilitsey of our developed code are then rigorously validated on three paradigmatic examples.
We use the extracted cubic and quartic IFCs to investigate anharmonic renormalization of phonon dispersions and thermal transport, with detailed comparison to previous calculations and experimental measurements.
In particular, bulk silicon, the most extensively studied semiconductor, is chosen to benchmark the basic functionalities of the \texttt{Pheasy} code;
we adopt monolayer tungsten disulfide (WS$_2$) as an instance to showcase the effectiveness of the code on studying two-dimensional (2D) materials, where the large mass difference between tungsten and sulfur atoms results in a considerable acoustic-optical gap and a strong four-phonon (4ph) scattering is thus expected;
cubic strontium titanate (SrTiO$_3$) is selected as an example for strongly anharmonic systems, and the SCP calculations are performed to obtain its stable phonon spectrum.
The computational details can be found in the ``Methods'' section.
We close this section with general guidelines for extracting high-order IFCs using supercell-based approaches.

\subsection*{Taylor expansion of the potential energy surface}

Within the Born--Oppenheimer approximation, the potential energy $V$ of a solid can be Taylor expanded in terms of ionic displacements $u_\mathbf{a}$ from their equilibrium positions $\mathcal{R}_\mathbf{a}$ as
\begin{equation}\label{eq:potential_taylor_expansion}
V=V_0+\Phi_\mathbf{a}u_\mathbf{a}+\frac{1}{2!}\Phi_\mathbf{ab}u_\mathbf{a}u_\mathbf{b}+\frac{1}{3!}\Phi_\mathbf{abc}u_\mathbf{a}u_\mathbf{b}u_\mathbf{c}+\cdots,
\end{equation}
where $V_0$ is a constant energy of the reference configuration and $\Phi_{\mathbf{a}_1\mathbf{a}_2\cdots\mathbf{a}_n}\equiv\partial^nV/\left( \partial u_{\mathbf{a}_1}\partial u_{\mathbf{a}_2}\cdots\partial u_{\mathbf{a}_n} \right)$ is the generic $n$th-order IFCs.
We use the bold letter $\mathbf{a}\equiv\{a,i\}$ to denote both an atom site $a$ in the lattice and a Cartesian component $i$, and the Einstein summation convention applies to repeated indices unless stated otherwise.
When the reference structure is at equilibrium, the linear expansion coefficient $\Phi_\mathbf{a}$ vanishes, i.e. no net forces on each atom.
By doing the first-order derivative of Eq.~\eqref{eq:potential_taylor_expansion} with respect to atomic displacements, a similar Taylor expansion for interatomic forces $F_\mathbf{a}$ is obtained as
\begin{equation}\label{eq:force_taylor_expansion}
F_\mathbf{a}\equiv-\frac{\partial V}{\partial u_\mathbf{a}}=-\Phi_\mathbf{ab}u_\mathbf{b}-\frac{1}{2}\Phi_\mathbf{abc}u_\mathbf{b}u_\mathbf{c}-\cdots.
\end{equation}
We then follow the multi-index notation introduced in the original work of CSLD~\cite{Zhou2019a} to rewrite the Taylor expansions of potential energy and interatomic forces into a more convenient cluster expansion form:
\begin{align}
V&=\frac{1}{\alpha!}\Phi_I(\alpha)u_I^\alpha,\label{eq:pot_cluster_expand} \\
F_\mathbf{a}&=-\frac{1}{\alpha!}\Phi_I(\alpha)\partial_\mathbf{a}u_I^\alpha,\label{eq:force_cluster_expand}
\end{align}
where $\alpha$ is the so-called \emph{cluster} consisting of $n$ lattice sites $\{a_1,a_2,\cdots,a_n\}$, $I\equiv\{ i_1,i_2,\cdots,i_n \}$ is a collection of Cartesian components, and $u_I^\alpha\equiv\prod_{n=1}^{|\alpha|}u_{a_n}i_n$ is the displacement polynomial.
For a given cluster $\alpha$, its absolute value and factorial are further defined as $|\alpha|\equiv\sum_a\alpha_a$ and $\alpha!\equiv\prod_a \alpha_a!$, respectively,
%
\iffalse
\begin{subequations}
\begin{align}
|\alpha|&\equiv\sum_a\alpha_a, \\
\alpha!&\equiv\prod_a \alpha_a!,
\end{align}
\end{subequations}
\fi
%
with $\alpha_a$ being the number of the repeated atom site $a$ within the cluster $\alpha$.
It is easy to notice that the absolute value of a cluster is simply equal to its length, which is the number of atom sites within it.
Importantly, the atoms in a cluster $\alpha$ can have an arbitrarily predefined order.
Since those clusters with the same sorted atom sites contribute equivalently to the lattice energy and interatomic forces, only one of them should be included in the expansions in Eqs.~\eqref{eq:pot_cluster_expand} and \eqref{eq:force_cluster_expand}.
In addition, for a cluster expansion~\cite{Sanchez1984}, we shall recall that a cluster with duplicate atom sites is termed \emph{improper}; otherwise, it is \emph{proper}.
Thanks to the locality of anharmonic IFCs, the expansion can be truncated by excluding those high-order clusters whose interatomic distances are larger than a specified cutoff radius, a common strategy used to reduce computational burdens.

\subsection*{Symmetry relations on force constants}

The expansion coefficients of a crystal potential energy in Eqs.~\eqref{eq:potential_taylor_expansion} and \eqref{eq:pot_cluster_expand} are not irreducible, and can transform to one another under the symmetry operations of that crystal catalog.
Since IFCs are the partial derivatives of crystal potentials, they must adhere to space-group symmetry, derivative commutativity and conservation of total crystal linear and angular momenta~\cite{Leibfried1961,Maradudin1968}.
Consequently, the number of independent IFC components are usually much smaller compared to their entire degrees of freedom, if those symmetry relations are properly accounted.
One advantage of using the prescribed cluster expansion notation is that it largely simplifies the symmetry analysis for IFCs.

\paragraph*{Space-group symmetry.} Atoms in crystals occupy lattice sites by conforming to the symmetries of the underlying space group $\mathbb{G}_S$, and IFCs thereby should transform covariantly in their constituent atoms of the corresponding clusters under the same symmetry operations.
Given any symmetry operation $\hat{s}\in\mathbb{G}_s$ transforming cluster $\alpha$ into $\hat{s}\alpha$, the following transformation rule holds for the IFCs of cluster $\alpha$~\cite{Zhou2019a}:
\begin{equation}\label{eq:spg_fc_transform}
\Phi_I(\hat{s}\alpha)=\Gamma_{IJ}(\hat{s})\Phi_J(\alpha)
\end{equation}
where $\Gamma_{IJ}(\hat{s})=\gamma_{i_1j_1}\gamma_{i_2j_2}\cdots\gamma_{i_nj_n}$ is a $3^n\times3^n$ matrix and $n$ is the order of IFCs (or cluster).
Specifically, in Cartesian coordinates, each symmetry operation $\hat{s}$ is comprised of an orthogonal transformation with a $3\times3$ unitary rotation $\boldsymbol{\gamma}$ followed by a translation $\mathbf{t}$, i.e. $\hat{s}\mathcal{R}_\mathbf{a}\equiv\boldsymbol{\gamma}\mathcal{R}_\mathbf{a}+\mathbf{t}$.
It is noteworthy that the atoms in mapped cluster $\hat{s}\alpha$ is generally ranked in an arbitrary order, depending on the detailed operation $\hat{s}$ applied, and an additional permutation $P_{II'}(\alpha)$ is hence required to obtain the transformed IFCs with atoms arranged in a specific order.

With the help of space-group symmetry, all possible clusters of lattice sites can be classified and grouped into the so-called \emph{orbits} which only contain the symmetry-equivalent clusters.
Given a cluster $\alpha$, its orbit is then defined as $G_{S\alpha}\equiv\{\hat{s}\alpha|\hat{s}\in\mathbb{G}_S\}$, which can be built by applying ergodically the operations in a space group.
Therefore, performing the cluster expansion of lattice potential in Eq.~\eqref{eq:pot_cluster_expand} is equivalent to find all symmetry-distinct clusters under space-group symmetry, also referred as \emph{representative} clusters.
The set of such representative clusters forms the entire \emph{orbit space} of cluster expansion.
In this sense, space-group symmetry operations can be divided into two subgroups: one mapping a cluster to its all symmetry-equivalent clusters (i.e. an orbit), and the other leaving the cluster invariant.
The latter is further denoted as the \emph{isotropy group} $\mathbb{G}_\alpha$ of cluster $\alpha$, which is a set of symmetry transformations leaving $\alpha$ invariant: $\mathbb{G}_\alpha\equiv\{\hat{s}\in\mathbb{G}_S|\hat{s}\alpha=\alpha\}$.
As a consequence, the cluster expansion of potential energy and interatomic forces in Eqs.~\eqref{eq:pot_cluster_expand} and \eqref{eq:force_cluster_expand} can be now recast into a form using only representative clusters~\cite{Zhou2019a}:
\begin{align}
V&=\sum_{\alpha\in\mathbb{A}/\mathbb{G}_S}\frac{1}{\alpha!}\sum_{\hat{s}\alpha\in\mathbb{G}_{S\alpha}}\Gamma_{IJ}(\hat{s})\Phi_J(\alpha)u_I^{\hat{s}\alpha},\label{eq:pot_cluster_expand_rep} \\
F_\mathbf{a}&=-\sum_{\alpha\in\mathbb{A}/\mathbb{G}_S}\frac{1}{\alpha!}\sum_{\hat{s}\alpha\in\mathbb{G}_{S\alpha}}\Gamma_{IJ}(\hat{s})\Phi_J(\alpha)\partial_\mathbf{a}u_I^{\hat{s}\alpha},\label{eq:force_cluster_expand_rep}
\end{align}
where $\mathbb{A}/\mathbb{G}_S$ stands for the orbit space with $\mathbb{A}$ denoting a set of all possible clusters.
In other words, using only representative clusters, the lattice sum can be decomposed into a sum over the clusters in the defined orbit space and a sum over the orbits corresponding to each representative cluster.
To simplify code implementation, the first atom of each representative cluster is selected within the reference unit cell, leveraging the periodicity of crystal lattice.
Finally, the space-group symmetry constraints on the IFCs of each representative cluster $\alpha$ can be constructed using the operations in its isotropy group $\mathbb{G}_\alpha$ as
\begin{equation}\label{eq:iso_const}
\Phi_I(\alpha)=\Gamma_{IJ}(\hat{s})\Phi_J(\alpha),
\end{equation}
Although the symmetry operations from isotropy group leave the cluster invariant, the order of atoms after transformation is not guaranteed to be unaltered.
In such cases, an additional permutation matrix $P_{II'}(\alpha)$ needs to be introduced in Eq.~\eqref{eq:iso_const} to ensure the same ordering before and after transformation, which is important to build the correct symmetry constraints on IFCs.

\paragraph*{Permutation symmetry.}

As IFCs are defined as the partial derivatives of lattice potential energy to atomic displacements, they must be commutative with respect to the order of differentiation.
Consequently, IFCs are invariant under the simultaneous permutations of atomic indices and Cartesian components as
\begin{equation}\label{eq:permute_const}
\Phi_I(\alpha)=P_{IJ}(\alpha)\Phi_J(\alpha),
\end{equation}
where $P_{IJ}(\alpha)$ is a $3^n\times3^n$ permutation matrix for a $n$th-order cluster $\alpha$, and we have used the same cluster notation $\alpha$ to represent the reordering of atoms in the permuted cluster for clarity.
Permutation symmetry needs to be imposed only on the IFCs of improper clusters over their duplicated atom sites, because the permutation relations of those involving improper clusters over different atoms or proper clusters will be handled directly during the mapping of representative clusters to their orbits.

\paragraph*{Global invariance conditions.} According to the well-known Noether's theorem~\cite{Noether1971}, the conservation of total linear and angular momenta in crystals requires IFCs to fulfill additional global invariances, i.e. the \emph{acoustic sum rules}.
For example, a rigid translation of the whole lattice by a constant displacement will result in the vanishing interatomic forces in Eq.~\eqref{eq:force_taylor_expansion}, which indicates the following translational invariance on generic IFCs:
\begin{equation}\label{eq:ti_const}
\sum_{\alpha\in\mathbb{A}/\mathbb{G}_S}\sum_{\hat{s}\alpha\in\mathbb{G}_{S\alpha}}^*\Gamma_{IJ}(\hat{s})\Phi_J(\alpha)=0,
\end{equation}
where the asterisk on the second summation means that only the clusters involving certain atomic indices in the orbits are summed, and each translational acoustic sum rule is identified with such unique array of atomic indices in length $n-1$.
Given $N_\mathrm{p}$ primitive unit cells with $N_\mathrm{a}$ atoms in each cell, there are in general $N_\mathrm{a}(N_\mathrm{a}N_\mathrm{p})^{n-2}$ matrix equations for the IFC order $n\geq2$, which can be further reduced dramatically to the number of representative clusters of the order $n-1$ by exploiting space-group symmetry.
Therefore, for $n$th-order IFCs, we only need to construct the translational invariance constraints identified by the atomic indices of $(n-1)$th-order representative clusters.

In addition, as a result of the global conservation of crystal angular momentum, rotational invariance conditions can be derived by adding an infinitesimal arbitrary rotation to the whole lattice~\cite{Leibfried1961}.
Such rotation matrix itself is of antisymmetric nature, since the symmetric part contributing to the real shear deformation of crystals must vanish.
Generally, rotational acoustic sum rules link the $(n+1)$th-order derivatives to the $n$th-order ones of lattice potential~\cite{Leibfried1961}, whose direct constructions will lead to an enormously large matrix for high-order IFCs.
Current implementations in \texttt{Pheasy} only adopt the lowest-order constraints on harmonic IFCs, i.e. the Born--Huang rotational invariance~\cite{Born1954}, and see Ref.~\cite{Lin2022} for details.
Besides, we further enforce the \emph{equilibrium conditions}~\cite{Born1954,Lin2022} for vanishing external stress on second-order IFCs.
Together with the Born--Huang rotational invariance, they are crucial for recovering the physically quadratic dispersion of ZA modes in the long-wavelegnth limit.
Notably, in infrared-active solids with the non-vanishing Born effective charges, the long-range electrostatic interactions also contribute to lattice potential and couple with external stress fields.
A separate treatment for long-range Coulomb interactions is thus needed, and we derived the so-called \emph{polar} invariance conditions~\cite{Lin2022} imposed on infrared-active solids.
Just as translational invariance guarantees the vanishing frequencies of three acoustic modes at the Brillouin zone center, rotational invariance is also important for the emergence of the fourth twisting acoustic mode in one-dimensional materials~\cite{Lin2022,Mounet2005}.

\subsection*{Force constant extraction as a linear problem}

To calculate IFCs, it is advantageous to use force-displacement relation in Eq.~\eqref{eq:force_cluster_expand_rep}, as each displaced configuration is able to provide $3N_\mathrm{a}N_\mathrm{p}-3$ independent force components.
For the case where force calculations are not accessible, e.g. the usage of certain exchange-correlation functionals, the energy-displacement relation given by Eq.~\eqref{eq:pot_cluster_expand_rep} is an alternative to extract IFCs.
Linear regression solvers are directly applicable after rewriting Eqs.~\eqref{eq:force_cluster_expand_rep} and \eqref{eq:pot_cluster_expand_rep} into a matrix-vector form as
\begin{align}
\mathbf{F}&=\mathbb{U}\cdot\boldsymbol{\Phi}=\mathbb{U}\cdot\mathbb{N}\cdot\boldsymbol{\phi}=\mathbb{W}\cdot\boldsymbol{\phi},\label{eq:force_mat_form} \\
\mathbf{V}&=\mathbb{U}^\mathrm{V}\cdot\boldsymbol{\Phi}=\mathbb{U}^\mathrm{V}\cdot\mathbb{N}\cdot\boldsymbol{\phi}=\mathbb{W}^\mathrm{V}\cdot\boldsymbol{\phi},\label{eq:pot_mat_form}
\end{align}
where $\boldsymbol{\Phi}$ is a flattened vector of the generalized IFC tensor of all representative clusters used in the expansion, $\mathbb{N}$ is the null space used to represent the irreducible components of IFCs, $\mathbb{W}=\mathbb{U}\cdot\mathbb{N}$ is the correlation (or sensing) matrix, and $\boldsymbol{\phi}=\mathbb{N}^{-1}\cdot\boldsymbol{\Phi}$ is a vector of independent IFC parameters to be solved; those variables with a superscript ``V'' are the counterparts for potential-energy fitting.
Explicitly, the two displacement polynomial $\mathbb{U}$ and $\mathbb{U}^\mathrm{V}$ have the shape of $N_F\times N_\Phi$, whose matrix elements take the following form~\cite{Zhou2019a}:
\begingroup
\allowdisplaybreaks
\begin{subequations}\label{eq:disp_mat}
\begin{align}
\mathbb{U}(\mathbf{a},\alpha I)&=-\frac{1}{\alpha!}\sum_{\hat{s}\alpha\in\mathbb{G}_{S\alpha}}\Gamma_{JI}(\hat{s})\partial_\mathbf{a}u_J^{\hat{s}\alpha},\label{eq:disp_mat_force}\\
\mathbb{U}^\mathrm{V}(\{u\},\alpha I)&=\frac{1}{\alpha!}\sum_{\hat{s}\alpha\in\mathbb{G}_{S\alpha}}\Gamma_{JI}(\hat{s})u_J^{\hat{s}\alpha},\label{eq:disp_mat_pot}
\end{align}
\end{subequations}
\endgroup
with $N_F$ and $N_\Phi$ as the total number of force and representative IFC components, respectively.

In order to precisely determine the number of independent IFC parameters of a given expansion, the null space $\mathbb{N}$ must be calculated exactly without any numerical approximation; it is constructed based on all kinds of IFC symmetry constraints [see Eqs.~\eqref{eq:iso_const} to \eqref{eq:ti_const}, etc.], which serves as a basis set to expand the irreducible components of IFC tensors.
We find that the iterative null space construction method presented in Ref.~\cite{Zhou2019a} is not accurate for low-symmetry crystals, whose null-space precision significantly depends on the selected tolerance to remove numerical instability.
Such method even completely fails when directly including rotational variance and equilibrium conditions into null space construction.
An extremely high precision of $\mathbb{N}$ is required in these cases, because the constraint equations of rotational variance and equilibrium conditions contain the Cartesian positions of atom sites that are irrational numbers.
One drawback of the column-pivot Gauss--Jordan elimination used in Ref.~\cite{Zhou2019a} is that the obtained row echelon form becomes numerically unstable, if the distribution of large elements in constraint matrix is highly uneven.
Therefore, to overcome the aforementioned difficulties, we have proposed an iterative maximal-pivot Gauss--Jordan elimination whose constructed $\mathbb{N}$ is exact and does not explicitly rely on the chosen tolerance.
The resulting null space $\mathbb{N}$ has the dimension of $N_\Phi\times N_\phi$, where $N_\phi$ is the number of irreducible IFC parameters.

In the \texttt{Pheasy} code, we mainly employ three solvers to solve the linear-regression problem in Eqs.~\eqref{eq:force_mat_form} and \eqref{eq:pot_mat_form}: the ordinary least square (OLS), least absolute shrinkage and selection operator (LASSO)~\cite{Tibshirani1996} and adaptive-LASSO (ALASSO)~\cite{Zou2006}.
Mathematically, their solutions are determined by minimizing the following respective cost functions:
\begin{subequations}\label{eq:linear_solver}
\begin{align}
\boldsymbol{\phi}^\mathrm{OLS}&=\underset{\boldsymbol{\phi}}{\arg\min}\frac{1}{2N_F}\|\mathbf{F}-\mathbb{W\cdot\boldsymbol{\phi}}\|^2_2,\label{eq:ols}\\
\boldsymbol{\phi}^\mathrm{LASSO}&=\underset{\boldsymbol{\phi}}{\arg\min}\frac{1}{2N_F}\|\mathbf{F}-\mathbb{W\cdot\boldsymbol{\phi}}\|^2_2+\mu\|\boldsymbol{\phi}\|_1,\label{eq:lasso}\\
\boldsymbol{\phi}^\mathrm{ALASSO}&=\underset{\boldsymbol{\phi}}{\arg\min}\frac{1}{2N_F}\|\mathbf{F}-\mathbb{W\cdot\boldsymbol{\phi}}\|^2_2+\mu w_i|\phi_i|,\label{eq:alasso}
\end{align}
\end{subequations}
where $\mu$ is a single hyperparameter used in both LASSO and adaptive LASSO, and $w_i$ are the feature-dependent weights in the adaptive LASSO.
While the OLS finds the optimal $\boldsymbol{\phi}$ solution through the minimization of Euclidian norm (i.e. $\ell_2$ norm $\|\cdots\|_2$), the LASSO algorithm introduces an additional penalty term using the Manhattan norm (i.e. $\ell_1$ norm $\|\cdots\|_1$).
Thanks to the combination of $\ell_1$ and $\ell_2$ regularization, LASSO can automatically pick a small subset of the physical relevant parameters out of the entire feature space, which has proven effective for underdetermined linear systems.
The sparsity of solution is controlled by the single hyperparameter $\mu$: small values of $\mu$ tends to yield highly sparse and underfitted results with only several non-zero IFCs, whereas large values of $\mu$ leads to an OLS-like fitting that is more susceptible to overfitting.
Thus, selecting an appropriate $\mu$ is a trade-off between the sparsity and quality of the fit, which is usually determined by a cross-validation scheme, to maximize the predictability of the obtained IFC model.
LASSO is widely applied in many scientific and engineering fields~\cite{Hastie2015}, particularly the success in CSLD~\cite{Zhou2014,Zhou2019a,Zhou2019b}; however, its feature selection has been shown to be biased, which is only consistent under certain conditions, and the standard LASSO procedure does not enjoy the general oracle properties~\cite{Zou2006}.
The feature-selection bias of a LASSO estimator can be overcome by using the adaptive LASSO~\cite{Zou2006}, where the $\ell_1$ regularization term is now accompanied with feature-dependent weights $w_i$ in Eq.~\eqref{eq:alasso}.
In general, the weights $w_i$ should be updated during LASSO iterations and reach the self-consistency, whose initial values are often guessed as the inverse of OLS solution, $w_i=1/\left|\phi^\mathrm{OLS}_i\right|$.
Consequently, in some situations, the computational cost of adaptive LASSO can become significantly expensive than standard LASSO, due to the additional convergence loop of finding optimal $w_i$.
The good one-shot approximation to $w_i$ is still the inverse of OLS solution, and in many cases we find that adaptive LASSO offers much sparser IFCs with a slightly smaller validation error.

For both LASSO and adaptive LASSO regressions, it is important to preprocess the sensing matrix $\mathbb{W}$ (or $\mathbb{W}^\mathrm{V}$) into a ``consistent'' manner across different data sets, since IFCs at different orders own different units and the $\ell_1$ penalty term depends on the magnitude of each IFC parameter.
In the original work of CSLD~\cite{Zhou2019a}, the authors used dimensionless displacements to construct sensing matrix by scaling with a ``maximum'' value which is selected on the order of thermal amplitude of lattice vibrations.
We here choose to perform the general standardization that is often used in machine learning, where each column of sensing matrix is standardized to have a vanishing mean value and a unity standard deviation.
Furthermore, it is noteworthy that both LASSO and adaptive LASSO algorithms perform quite well on an underdetermined sensing matrix, while the sensing matrix in OLS fitting must be full-rank.
Given an overdetermined linear system, finding the OLS solution is equivalent to perform the Moore--Penrose pseudoinverse~\cite{Golub2013}, i.e. $\boldsymbol{\phi}^\mathrm{OLS}=\mathbb{W}^+\cdot\mathbf{F}$ with the pseudoinverse $\mathbb{W}^+=(\mathbb{W}^\top\cdot\mathbb{W})^{-1}\cdot\mathbb{W}^\top$ .
When randomly displaced supercells are used, a good estimation for the minimum number of configurations to fulfill the full-rank condition is $\sim$$N_\phi/(3N_\mathrm{a}N_\mathrm{p})$, and it is much safer to increase it by a fold of 2 to 5 in the OLS case to mitigate the issue of overfitting.
The \texttt{Pheasy} code directly utilizes the LASSO implementation from the \texttt{scikit-learn} package~\cite{Pedregosa2011}, and further implements the adaptive LASSO that is not available there.

\subsection*{Generation of training structures}
The selection and preparation of training dataset essentially impacts both IFC training efficiency and the accuracy of the resulting lattice model.
Good training structures not only reduce the number of required displaced configurations, but also ensure the robustness and fidelity of the extracted IFCs.
Ideally, they should represent the most physically favored landscape of the potential energy surface and be mutually independent to minimize possible correlations in the sensing matrix $\mathbb{W}$.
In compressive-sensing signal recovery, this corresponds to the usage of random samples that are independent and identically distributed~\cite{Candes2008}.
The $\texttt{Pheasy}$ code offers three common recipes to generate highly effective training configurations satisfying the aforementioned requirements: (i) random displacements of fixed magnitude, (ii) populations from quantum canonical sampling of normal modes at a given temperature, and (iii) \emph{ab initio} molecular dynamics (AIMD) trajectories.

In the first scenario, training structures are generated by moving atoms along a random direction with a constant displacement magnitude, and all atoms in the supercell are displaced.
This strategy is particularly useful for extracting only harmonic IFCs or up to cubic anharmonicity, where we find that in most cases the constant random displacements of 0.01 and 0.03 \r{A} are good choices for second- and third-order IFCs, respectively.
Second, since ionic probability distribution is well approximated using a Gaussian function~\cite{Monacelli2021}, it is appealing to sample training structures from the populations of a quantum canonical ensemble for harmonic oscillators at a given temperature.
The covariance matrix $\boldsymbol{\Psi}$ entering such a Gaussian distribution is the displacement-displacement correlation function $\Braket{u_\mathbf{a}u_\mathbf{b}}$, which can be calculated within the harmonic approximation as~\cite{Monacelli2021}
\begin{equation}\label{eq:covar_mat}
\Psi_\mathbf{ab}=\Braket{u_\mathbf{a}u_\mathbf{b}}=\frac{\hbar}{2\sqrt{m_am_b}}\sum_\nu\frac{2n_\nu+1}{\omega_\nu}e_{\nu\mathbf{a}}e_{\nu\mathbf{b}},
\end{equation}
where $\Braket{\cdots}$ denotes an ensemble average at a given temperature, $\hbar$ is the reduced Planck constant, $m_a$ is the atomic mass, and $n_\nu$ is the Bose--Einstein distribution of the phonon mode $\nu$ with its frequency $\omega_\nu$ and eigenvector $e_{\nu\mathbf{a}}$.
Here, Eq.~\eqref{eq:covar_mat} is evaluated directly based on a supercell using only the $\Gamma$ point, and it is fully equivalent to sample a phonon grid of primitive unit cell in reciprocal space that is commensurate with the supercell dimension.
The calculation of $\boldsymbol{\Psi}$ needs to know the phonon spectrum of system, which is usually obtained based on the initial harmonic IFCs from either the small displacement method~\cite{Alf2009,Togo2015,Togo2023a,Togo2023b}, DFPT~\cite{Baroni2001,Gonze1997a} or fixed-magnitude random displacement approach.
Then, displaced configurations are stochastically depicted using a multivariate Gaussian $\rho(\{u\})=\sqrt{\det(\boldsymbol{\Psi}^{-1}/2\pi)}\exp(-\frac{1}{2}u_\mathbf{a}\cdot\Psi^{-1}_\mathbf{ab}\cdot u_\mathbf{b})$.
The main advantages of this canonical sampling technique are that it yields the physically real displacements at a given temperature, and avoids the guess of displacement magnitude, which is difficult to know for the IFCs beyond third order.
Last, in the third case, one can also utilize the snapshots from AIMD trajectories, provided they are sampled at a sufficiently large interval to mitigate potential correlations among them.
Due to the high computational cost and lack of quantum effects, it is generally not recommended to use AIMD for the generation of training dataset.

\subsection*{Long-range electrostatic interactions}

Before moving to the pedagogical calculations using \texttt{Pheasy}, a potential caveat regarding the IFC extraction in infrared-active solids should be noted.
The microscopic interactions in infrared-active solids are inherently long-ranged, due to the spontaneous electric polarizations induced by thermal motions of ions.
Generally speaking, owing to insufficient electronic screening, any semiconductor or insulator also exhibits the long-range Coulomb interactions which are of dynamical multipolar nature.
An appropriate strategy is to decompose the total interatomic forces $F_\mathbf{a}$ into a short-range ($\mathcal{S}$) and long-range ($\mathcal{L}$) component as $F_\mathbf{a}=F^\mathcal{S}_\mathbf{a}+F^\mathcal{L}_\mathbf{a}$, where $F^\mathcal{L}_\mathbf{a}$ can be calculated analytically in reciprocal space followed by a Fourier transform into real space.
For bulk solids, the generic long-range contribution to dynamical matrix in the long-wavelength limit can be expressed as~\cite{Stengel2013}
\begin{equation}\label{eq:long-range}
\Phi_{\kappa i,\kappa'j}^{\mathcal{L}}(\mathbf{q})=\lim_{\mathbf{q}\to\mathbf{0}}\frac{4\pi}{\Omega}\frac{\overline{\rho}_{\kappa i}(\mathbf{q})\overline{\rho}_{\kappa' j}(\mathbf{q})}{\xi(\mathbf{q})},
\end{equation}
where $\overline{\rho}_{\kappa i}(\mathbf{q})$ is the unscreened bare charge induced by the collective excitation at the momentum $\mathbf{q}$ with the $\kappa$th atom in the primitive cell displaced along the Cartesian direction $i$, $\xi(\mathbf{q})$ is the dielectric screening function even in $\mathbf{q}$, and $\Omega$ is the volume of primitive unit cell.
Both $\overline{\rho}_{\kappa i}(\mathbf{q})$ and $\xi(\mathbf{q})$ are analytic in the long-wavelength limit ($\mathbf{q}\to\mathbf{0}$), which can be further expanded into a Taylor series~\cite{Stengel2013}.
In the lowest-order approximation, we have $\overline{\rho}_{\kappa i}(\mathbf{q})\approx-\imath eq_j Z^j_{\kappa i}/\Omega$ ($\imath$ the imaginary unit) and $\xi(\mathbf{q})\approx \boldsymbol{\epsilon}^{\infty}\cdot(\mathbf{q}\otimes\mathbf{q})$, and Eq.~\eqref{eq:long-range} becomes the well-known dipole-dipole (DD) interactions:
\begin{equation}\label{eq:DD_term}
\Phi_{\kappa i,\kappa'j}^{\mathcal{L}}(\mathbf{q})\approx\boldsymbol{\Phi}_{\kappa i,\kappa'j}^\mathrm{DD}(\mathbf{q})=\frac{4\pi e^2}{\Omega}\frac{\left(\mathbf{q}\cdot\mathbf{Z}_\kappa\right)_i\left(\mathbf{q}\cdot\mathbf{Z}_{\kappa'}\right)_j}{\mathbf{q}\cdot\boldsymbol{\epsilon}^{\infty}\cdot\mathbf{q}},
\end{equation}
where $\mathbf{Z}_\kappa$ is the second-rank dynamical Born effective charge of the $\kappa$th atom in the primitive cell, $\boldsymbol{\epsilon}^{\infty}$ is the clamped-ion dielectric permittivity tensor, and $e$ is the elementary charge.
It is easy to notice that dipole-dipole interactions are the zeroth order in $\mathbf{q}$, which is non-analytic at $\mathbf{q}=\mathbf{0}$, and the LO-TO splitting thus depends on the direction approaching the long-wavelength limit.
Thanks to our recent development~\cite{Lin2024}, the high-order expansion of Eq.~\eqref{eq:long-range} beyond the dipole approximation has also been available, where we are able to construct the long-range multipolar interactions up to the second-order in $\mathbf{q}$; these will further include the dynamical quadrupole- and octupole-related interactions.
Although we here focus on dipole-dipole interactions for clarity, the formulation of removing long-range Coulomb forces is general.
Specifically, the implementation in \texttt{Pheasy} has been already extended to contain high-order multipolar interactions and see Ref.~\cite{Lin2024} for details.
For practical calculations of Eq.~\eqref{eq:DD_term}, the Ewald summation technique~\cite{Baroni2001,Gonze1997a} is adopted in the code, and the implementation details can also be found in Ref.~\cite{Lin2024}.
Once the long-range dynamical matrix is calculated, the corresponding contribution to real-space IFCs can be constructed efficiently via Fourier transform as
\begin{equation}\label{eq:long_range_IFCs}
\Phi^\mathcal{L}_{\kappa i,\kappa'j}(\mathbf{R}=\mathbf{0},\mathbf{R}')=\frac{1}{N_\mathbf{q}}\sum_\mathbf{q}\Phi_{\kappa i,\kappa'j}^{\mathcal{L}}(\mathbf{q})\mathrm{e}^{-\imath \mathbf{q}\cdot\mathbf{R}'},
\end{equation}
where $N_\mathbf{q}=N_\mathrm{p}$ is the number of $\mathbf{q}$-points sampled in reciprocal space (equal to the number of primitive unit cells in real space), $\mathbf{R}$ is the position of unit cell.
We have set the reference unit cell to $\mathbf{R}=\mathbf{0}$ by exploiting the translational invariance of lattice.

Finally, the short-range interatomic forces after removing long-range contributions can be readily obtained as $F_\mathbf{a}^\mathcal{S}=F_\mathbf{a}-F_\mathbf{a}^\mathcal{L}=F_\mathbf{a}+\Phi_\mathbf{ab}^\mathcal{L}u_\mathbf{b}$.
Therefore, in infrared-active solids, $F_\mathbf{a}^\mathcal{S}$ should be used in Eq.~\eqref{eq:force_mat_form} to ensure the sufficient decay of IFCs in real space, a \emph{priori} of CSLD.
The removal of dipole-dipole interactions is able to guarantee the leading harmonic IFCs decaying faster than $1/d^3$~\cite{Gonze1997a}, and a faster-than-quartic decay requires the further elimination of high-order multipolar interactions~\cite{Stengel2013}.
The extracted second-order IFCs are now short-ranged, and the long-range Coulomb interactions will be added back later using the same analytic formula in Eq.~\eqref{eq:DD_term} when performing phonon calculations.
In addition, we note that the exact form of screened Coulomb kernel depends on system dimensionality, and the 2D version of long-range electrostatic interactions has also been implemented in the \texttt{Pheasy} code following Refs.~\cite{Sohier2017,Royo2021}.
Interestingly, through our tests, we find that infrared-active materials do not necessarily require the prescribed scheme above; the IFC extractions using LASSO with and without removing long-range components show a similar regression performance as well as almost identical phonon properties.
This can be understood that long-range contributions only enter harmonic IFCs that are generally kept as dense tensors, and hence the sparsity of anharmonic IFCs should remain unaffected.
Such observation contrasts with the conclusion drawn by Zhou et al.~\cite{Zhou2019b}, and the role of long-range Coulomb interactions in IFC extractions deserves a future investigation.

\begin{figure*}[t!]
\centering
\includegraphics[width=0.99\linewidth]{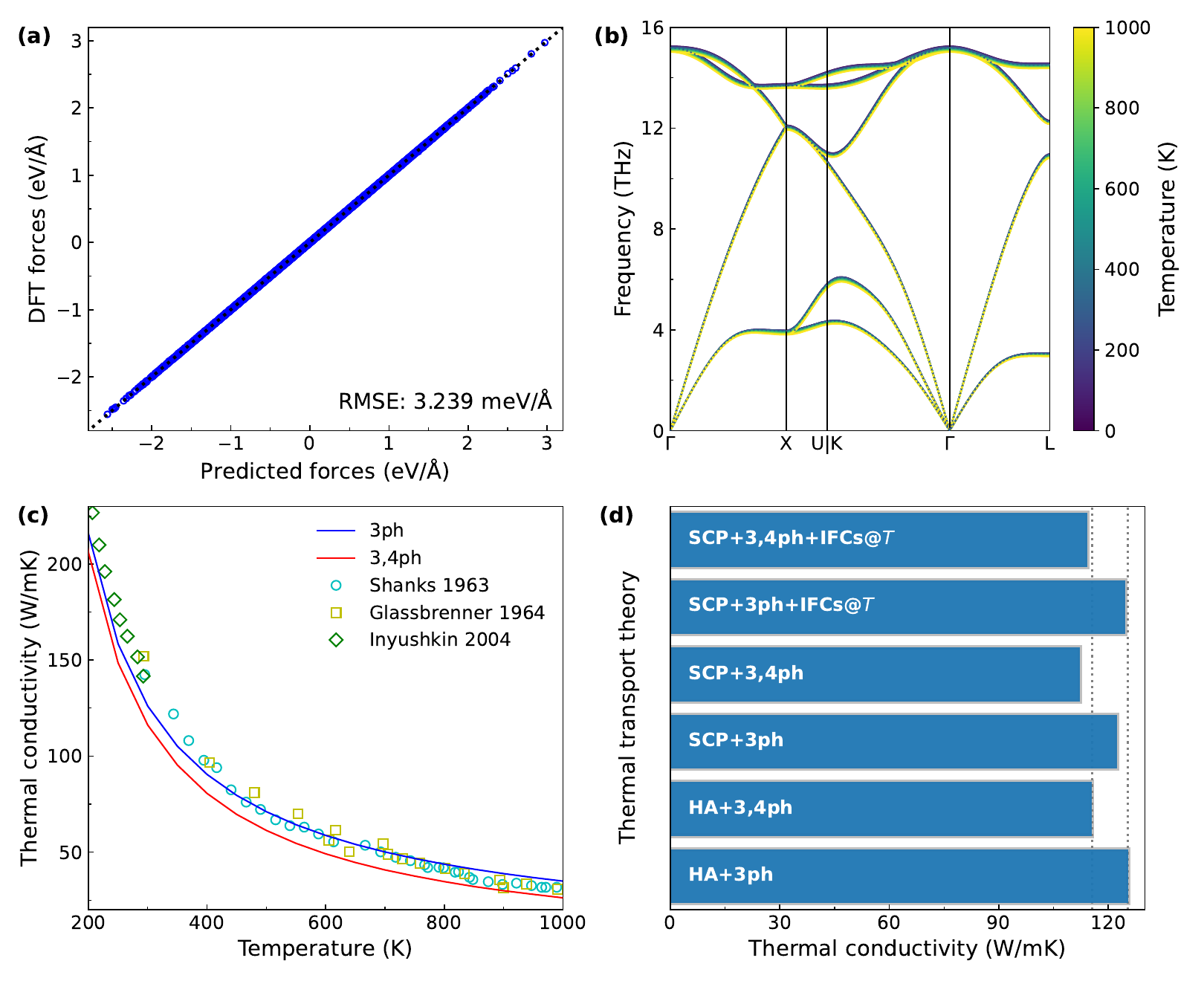}
\caption[Anharmonic lattice dynamics and thermal transport in bulk silicon]{\textbf{Anharmonic lattice dynamics and thermal transport in bulk silicon:}
(a) verification of the trained IFC model for predicting interatomic forces against DFT results;
(b) temperature-dependent phonon dispersions from 0 to 1000~K using SCP calculations with the loop diagram of phonon self-energy, where the room-temperature one is depicted in dotted lines;
(c) LTC as a function of temperature from 200 to 1000~K, compared with experimental measurements~\cite{Shanks1963,Glassbrenner1964,Inyushkin2004};
(d) room-temperature LTC calculated with the increasing level of approximations in thermal transport theory, where ``HA'', ``3,4ph'' and ``IFCs@$T$'' denote the harmonic approximation, three- and four-phonon scattering, and temperature-dependent anharmonic IFCs, respectively.
The two vertical dash lines in panel (d) are used to guide the eyes, which represent the thermal conductivities at the ``HA+3ph'' and ``HA+3,4ph'' levels, which are also the case of the LTC in panel (c).
}
\label{fig:silicon}
\end{figure*}

\subsection*{Benchmark: bulk silicon}
We first benchmark the accuracy and reliability of the \texttt{Pheasy} code on bulk silicon, the most commercially used semiconductor.
Its thermal transport properties are widely investigated by both theoretical calculations~\cite{Li2014,Han2022,Broido2007,Gu2020,Kim2020,Guo2024,Castellano2025} and experimental measurements~\cite{Shanks1963,Glassbrenner1964,Inyushkin2004}, including the impact of temperature effects on the phonon spectrum.
Using a $4\times4\times4$ supercell (128 atoms), we compute the initial harmonic IFCs with 5 randomly displaced configurations, where all atoms are moved in random directions with a fixed magnitude of 0.01~\r{A}.
Then, the training dataset for extracting IFCs up to sixth-order is randomly generated according to the populations of a quantum canonical ensemble at 300~K, following the Gaussian distribution characterized by the covariance matrix in Eq.~\eqref{eq:covar_mat}.
Each ensemble is a training dataset and contains 64 displaced structures.
Since anharmonic IFCs are quite localized in real space, we consider a fifth-nearest neighbor ($\sim$6.3~\r{A}) as the cutoff distance for the third-order IFCs and a third-nearest neighbor ($\sim$5.0~\r{A}) for all other higher-order ones.
We further restrict the fifth- and sixth-order IFCs to have maximum three-body interactions, excluding their tensors involving more than three different atoms in the expansion.
A previous study~\cite{Tadano2015} has demonstrated the effectiveness of this trick in reducing the total number of irreducible IFC components, without affecting the accuracy of the Taylor-expanded lattice model.
By doing such a expansion of lattice potential using a $4\times4\times4$ supercell of silicon, we have totally 2239 independent IFC parameters, with 44, 199, 581, 474 and 941 components for each order, respectively.
After collecting interatomic forces from DFT for a training ensemble, we fit these parameters using the adaptive LASSO.
As shown in Fig.~\ref{fig:silicon}(a), our constructed interatomic potential using a sixth-order Taylor expansion yields a high accuracy in predicting interatomic forces, in excellent agreement with direct DFT calculations.
The test dataset in Fig.~\ref{fig:silicon}(a) is another independent ensemble at 300~K using the aforementioned preparation scheme, containing  64 supercell structures.
Overall, the decent performance of our trained lattice model is corroborated by a very small root mean square error (RMSE) of 3.239~meV/\r{A}.

We proceed to investigate the temperature-dependent phonon dispersions using SCP calculations, considering only the lowest-order loop self-energy of quartic anharmonicity.
The self-consistent equations for anharmonic renormalization of second-order IFCs can be derived by taking the second-order derivative of the Taylor-expanded potential energy in Eq.~\eqref{eq:potential_taylor_expansion} with respect to atomic displacements and performing the canonical ensemble average at a given temperature:
\begin{equation}\label{eq:eff_2nd_IFCs}
\Phi^\mathrm{eff}_\mathbf{ab}=\Phi_\mathbf{ab}+\Phi_\mathbf{abc}\Braket{u_\mathbf{c}}+\frac{1}{2}\Phi_\mathbf{abcd}\Braket{u_\mathbf{c}u_\mathbf{d}}+\mathcal{O}(u^3),
\end{equation}
where $\Phi^\mathrm{eff}_\mathbf{ab}\equiv\Braket{\frac{\partial^2V}{\partial u_\mathbf{a}\partial u_\mathbf{b}}}$ are the temperature-dependent effective harmonic IFCs.
If neglecting the temperature-induced structure relaxation effects (i.e. the tadpole diagram of phonon self-energy~\cite{Maradudin1962,Paulatto2015}), the term involving cubic anharmonicity, $\Phi_\mathbf{abc}\Braket{u_\mathbf{c}}$, can be dropped, and we therefore have the final self-consistent equations as
\begin{equation}\label{eq:SCP_loop}
\Phi^\mathrm{eff}_\mathbf{ab}=\Phi_\mathbf{ab}+\frac{1}{2}\Phi_\mathbf{abcd}\Braket{u_\mathbf{c}u_\mathbf{d}},
\end{equation}
with $\Braket{u_\mathbf{c}u_\mathbf{d}}$ given by Eq.~\eqref{eq:covar_mat}.
By defining a displacement operator $u_\mathbf{a}=\sqrt{\frac{\hbar}{2m_a\omega_\nu}}B_\nu e_{\nu\mathbf{a}}$ in the second quantization~\cite{Maradudin1962,Dove1993}, the displacement-displacement response function $\Braket{u_\mathbf{a}u_\mathbf{b}}$ is shown to be proportional to phonon Green's function $\Braket{\mathcal{T}\left[B_\nu(t)B^\dagger_{\nu'}(t)\right]}$, in which $\mathcal{T}$ is the time ($t$)-ordering operator and $B_\nu=b_\nu+b^\dagger_\nu$ is the ladder operator with $b_\nu$ and $b^\dagger_\nu$ as the phonon annihilation and creation operators.
Based on this definition, one can demonstrate that Eq.~\eqref{eq:SCP_loop} is just the Fourier transform of the Dyson equation for a phonon propagator with the loop diagram~\cite{Tadano2015,Paulatto2015,Ravichandran2018}.
Since the loop self-energy is real, it is convenient and highly efficient to perform the anharmonic renormalization of second-order IFCs using Eq.~\eqref{eq:SCP_loop}, thereby termed as the real-space SCP approach.
While $\Phi^\mathrm{eff}_\mathbf{ab}$, $\Phi_\mathbf{ab}$ and $\Phi_\mathbf{abcd}$ are all given in real space, we compute $\Braket{u_\mathbf{a}u_\mathbf{b}}$ in reciprocal space with a phonon grid commensurate to the supercell dimension of IFCs.
An equivalent implementation is also provided by Ravichandran and Broido~\cite{Ravichandran2018}, which they called the statistical perturbation-operator renormalization.
The complete microscopic theory of our SCP approach, including both cubic and quartic anharmonicities, will be presented in a forthcoming paper.

Fig.~\ref{fig:silicon}(b) shows the temperature-dependent phonon dispersions of silicon from 0 to 1000~K, calculated using our real-space SCP approach.
As temperature increases, the phonon dispersions exhibit an overall redshift, and the effects on optical branches are more visible than acoustic ones.
Indeed, the temperature renormalization effects are almost negligible by considering only the loop diagram, and a more pronounced softening of the phonon spectrum in silicon was observed at high temperatures when the bubble diagram of cubic anharmonicity was included using the TDEP method~\cite{Gu2020,Kim2020}.
However, we should emphasize that it is incorrect to use phonon frequencies with bubble corrections in LTC calculations, which is because phonon-phonon scattering rates are evaluated perturbatively using Fermi's golden rule without achieving the self-consistency.
Since 3ph linewidths are the imaginary part of the bubble self-energy~\cite{Maradudin1962}, using bubble-corrected phonon frequencies without self-consistently updating the bubble diagram for 3ph scattering rates violates many-body perturbation theory.
Therefore, in standard LTC predictions using the Boltzmann transport equation, temperature-dependent phonon frequencies should be calculated with only the loop and tadpole diagrams.

\begin{figure*}[th!]
\centering
\includegraphics[width=0.99\linewidth]{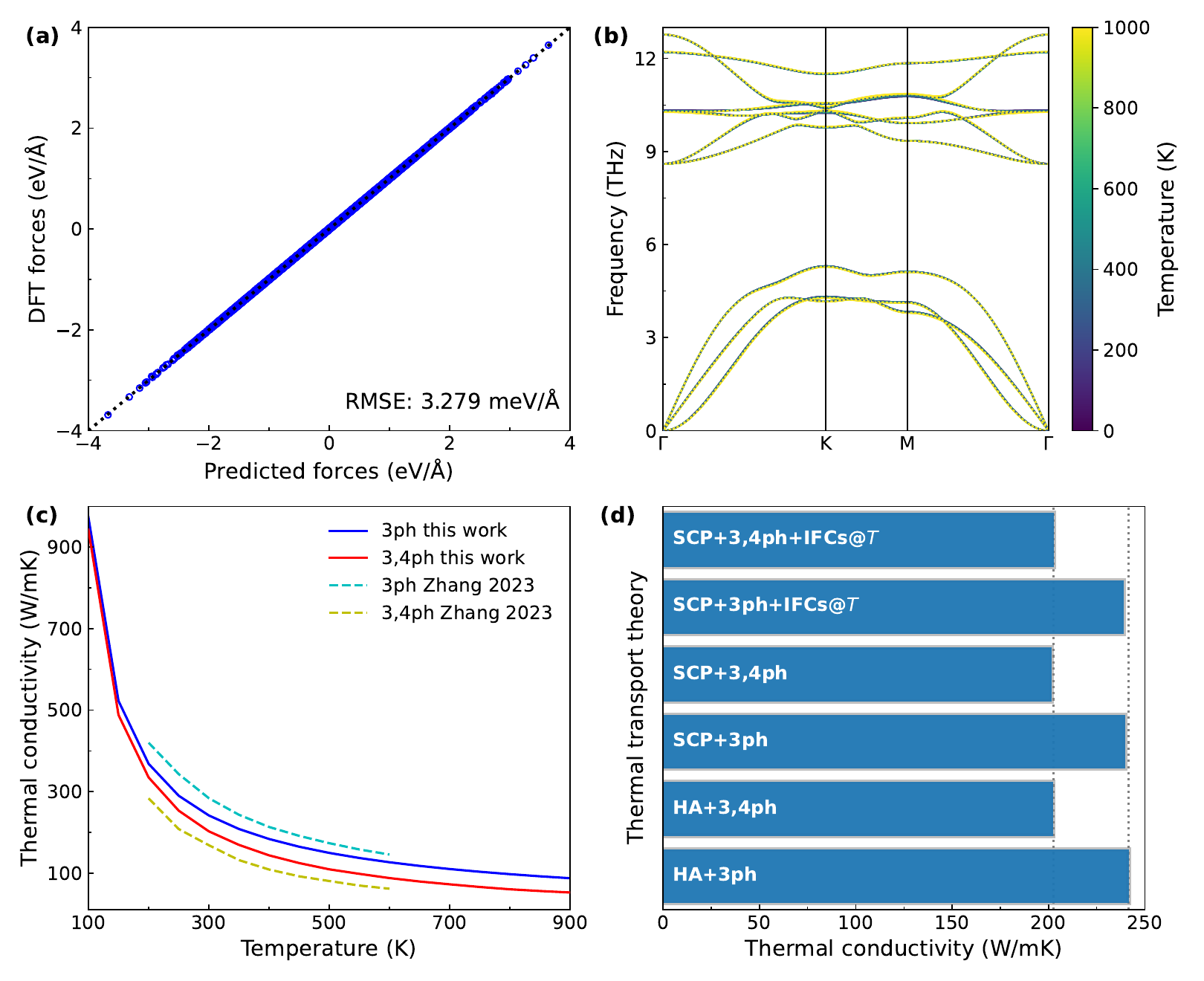}
\caption[Anharmonic lattice dynamics and thermal transport in monolayer WS\textsubscript{2}]{\textbf{Anharmonic lattice dynamics and thermal transport in monolayer WS\textsubscript{2}:}
(a) verification of the trained IFC model for predicting interatomic forces against DFT results;
(b) temperature-dependent phonon dispersions from 0 to 1000~K using SCP calculations with the loop diagram of phonon self-energy, where the room-temperature one is depicted in dotted lines;
(c) LTC as a function of temperature from 100 to 900~K, compared with previous theoretical calculations~\cite{Zhang2023};
(d) room-temperature LTC calculated with the increasing level of approximations in thermal transport theory.
The two vertical dash lines in panel (d) are used to guide the eyes, which represent the thermal conductivities at the ``HA+3ph'' and ``HA+3,4ph'' levels, which are also the case of the LTC in panel (c).
}
\label{fig:ws2}
\end{figure*}

The LTC of silicon as a function of temperatures is illustrated in Fig.~\ref{fig:silicon}(c), where the temperature dependence of phonon dispersions and anharmonic IFCs are neglected.
When only considering 3ph scattering, the calculated LTC shows an overall good agreement with its experimental results~\cite{Shanks1963,Glassbrenner1964,Inyushkin2004} in the entire temperature range.
Nonetheless, our simulations underpredict the LTC below 400~K, which should be attributed to the use of the PBEsol functional.
Our 3ph LTC of silicon at 300~K is 125.4~W/mK, in good agreement with two recent calculations using the same PBEsol functional, around 124~\cite{Castellano2025} and 120--130 W/mK~\cite{Seko2024}, respectively.
Particularly, using consistent DFT parameters, the room-temperature LTC of silicon calculated based on the second-/third-order IFCs from \texttt{Phonopy}/\texttt{thirdorder.py} is 124.3 W/mK, and it also agrees well with our results across the entire temperature range (see the Supplementary Note 2 and Fig.~S2 for details).
The inclusion of 4ph scattering further lowers the LTC of silicon, and the decrease becomes more pronounced at high temperatures.
Specifically, 4ph processes result in a 7.9\% drop at 300~K with a final LTC of 115.4 W/mK, and a recent calculation using the same PBEsol functional gave a value of 112 W/mK~\cite{Castellano2025}.
In addition, we provide in Fig.~\ref{fig:silicon}(d) the room-temperature LTC of silicon calculated with the increasing level of thermal transport theory.
The temperature-dependent anharmonic IFCs (denoted as ``IFCs@$T$'' in the plot) are obtained by fitting the effective third- and fourth-order IFCs of our randomly generated ensembles at 300 K with the harmonic IFCs fixed at the SCP solutions.
This procedure needs to achieve the self-consistency in the ensemble populations, as the initial ensemble is depicted using the second-order IFCs at the ground state.
We find that two ensembles are already enough to achieve the self-consistency in the LTC calculated with temperature-dependent anharmonic IFCs (see Table~1 of the Supplementary Information for more details), which also reveals that silicon is a weakly anharmonic solid.
Indeed, as shown in Fig.~\ref{fig:silicon}(d), employing different levels of approximations in thermal transport simulation does not lead to noticeable difference in the room-temperature LTC of silicon, where 3ph and 4ph scattering almost play a similar role in all cases.
In short, our first example on bulk silicon has already demonstrated the high accuracy and reliability of the \texttt{Pheasy} code in simulating phonon-related physical properties.

\subsection*{Application: tungsten disulfide monolayer}
Monolayer WS$_2$ is one of the promising 2D semiconductors, and a recent high-throughput calculation~\cite{Ha2024} unveiled its ultrahigh hole mobility at room temperature among other transition metal dichalcogenides.
Efficient heat dissipation is essential to maintain desirable performance of the fabricated electronic devices.
We now apply our code to study lattice dynamics and thermal transport in 2D WS$_2$.
Similar to the case of silicon, we prepare training and testing datasets using finite-temperature displacements that are randomly generated based on quantum canonical ensembles at 300 K, and each ensemble contains 64 supercell configurations.
DFPT calculations are performed to obtain the initial harmonic IFCs for a $6\times6\times1$ supercell, which are used to construct the displacement covariance matrix in Eq.~\eqref{eq:covar_mat} at 300 K.
A sixth-order cluster expansion is adopted to represent the potential energy surface of 2D WS$_2$, where we have truncated third-order and other higher-order IFCs with a cutoff distance of sixth- ($\sim$6.6~\r{A}) and fourth-nearest ($\sim$5.3~\r{A}) neighbors, respectively.
Besides, for quartic IFCs, we omit the negligible four-body interactions, and we consider only the onsite and two-body terms for fifth- and sixth-order IFCs.
After imposing symmetry constraints, the numbers of irreducible IFC components for the second to sixth orders are 140, 1194, 1342, 75 and 107, respectively.
Our IFC model is then trained by using the LASSO algorithm, whose accuracy and predictability is shown in Fig.~\ref{fig:ws2}(a).
By validating on an independent ensemble with 64 structures, we obtain a RMSE of 3.279 meV/\r{A} for  predicting interatomic force against DFT references.

We then use our SCP scheme in Eq.~\eqref{eq:SCP_loop} to investigate temperature-dependent phonon spectra of WS$_2$.
As shown in Fig.~\ref{fig:ws2}(b), there is no obvious temperature effect on the calculated phonon dispersions from 0 to 1000 K.
This observation is consistent with previous results from TDEP~\cite{Zhang2023} and spectal energy density analysis~\cite{Mobaraki2019}, which indicates that WS$_2$ monolayer is a weak anharmonic solid.
Fig.~\ref{fig:ws2}(c) presents the calculated temperature-dependent LTC of 2D WS$_2$, where the temperature effects on phonon dispersions or anharmonic IFCs are not considered.
We obtain a 3ph LTC of 241.6 W/mK at 300~K, in good agreement with a recent calculation ($\sim$284 W/mK) using IFCs from a machine-learning potential~\cite{Zhang2023}.
When 4ph collisions are activated, it leads to a reduction of 16.2\% and gives rise to a LTC of 202.5 W/mK at room temperature, whereas Zhang et al.~\cite{Zhang2023} found a more remarkable decrease of 34.5\% due to 4ph scattering.
Interestingly, as depicted in Fig.~\ref{fig:ws2}(c), their 3ph LTC is overall higher than our prediction; however, they have a lower LTC when 4ph scattering is also included.
Indeed, the LTC of 2D WS$_2$ also varied significantly in previous calculations, ranging from 142~\cite{Gu2014} to 299.87 W/mK~\cite{Han2021}.
The \emph{in situ} experimental measurements on such 2D monolayers remain elusive, which have produced a dramatically lower LTC, 63$\pm$7~\cite{Sang2022} and 32~W/mK~\cite{Peimyoo2014} at room temperature.
Therefore, the LTC of 2D WS$_2$ awaits more refined thermal measurements in the future.
Furthermore, we employ a hierarchy of thermal transport theory as in Fig.~\ref{fig:ws2}(d) to examine the LTC of 2D WS$_2$.
At 300~K, we find there is no evident difference in the calculated LTC, when the effects of temperature-dependent phonon spectra and anharmonic IFCs are taken into account.
This is consistent with Fig.~\ref{fig:ws2}(b), where phonon dispersions also exhibit negligible temperature effects.
Specifically, the temperature-dependent anharmonic IFCs converge after two ensemble iterations and see Table~1 of the Supplementary Information for the LTC calculated in each ensemble.
All these observations confirm that 2D WS$_2$ is of weak anharmonicity.
Not least, it is worthy noting that Zhang et al.~\cite{Zhang2023} found a prominent effect in the LTC using temperature-dependent anharmonic IFCs, yielding a 3,4ph value of $\sim$214 W/mK at 300 K, which is closer to our results.
Such difference could be attributed to the overshoot anharmonicity in the TDEP method~\cite{Monacelli2025}.

\begin{figure*}[t!]
\centering
\includegraphics[width=0.99\linewidth]{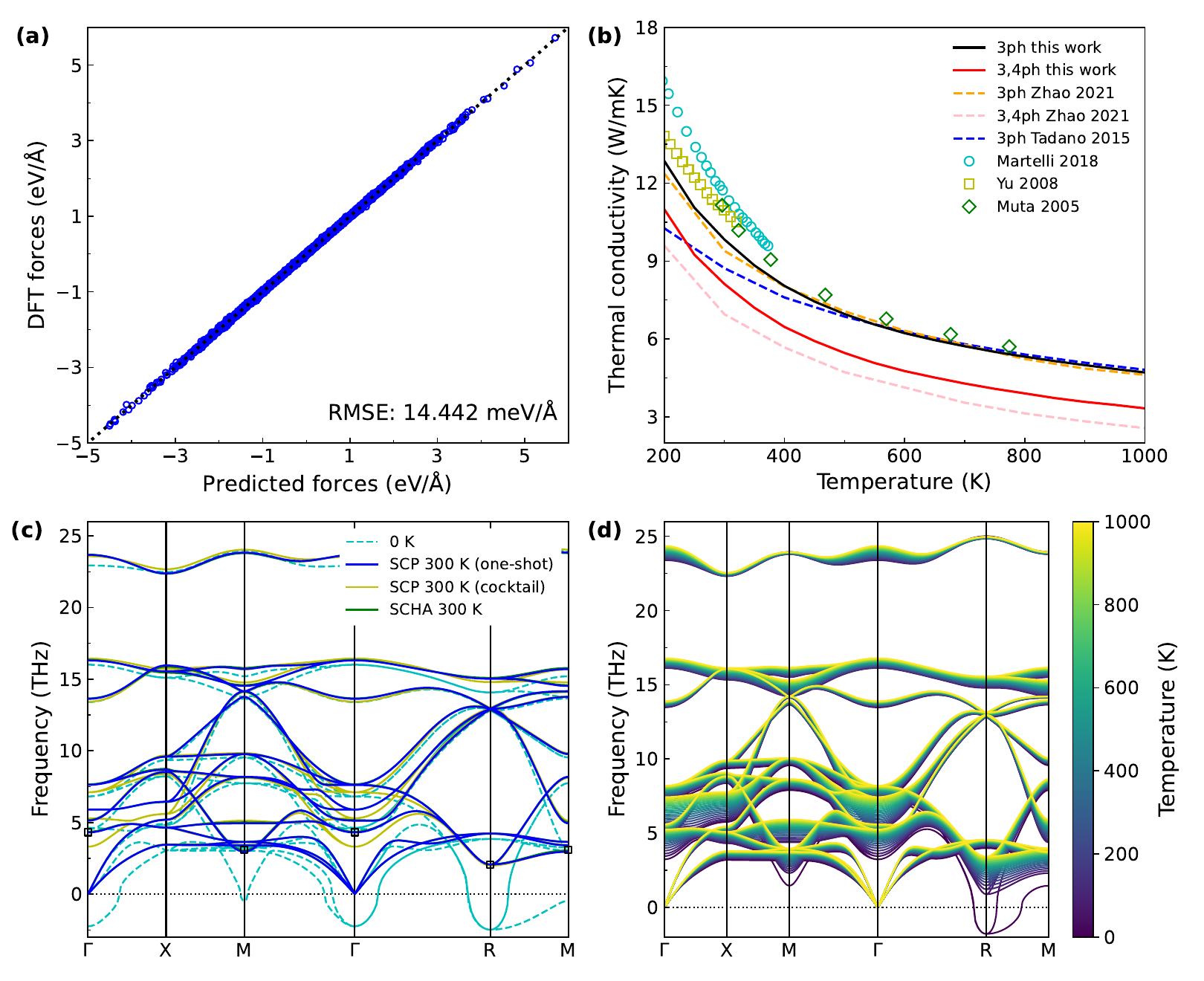}
\caption[Anharmonic lattice dynamics and thermal transport in cubic SrTiO\textsubscript{3}]{\textbf{Anharmonic lattice dynamics and thermal transport in monolayer SrTiO\textsubscript{3}:}
(a) verification of the trained IFC model for predicting interatomic forces against DFT results;
(b) LTC as a function of temperature from 200 to 1000~K, compared with other theoretical calculations~\cite{Tadano2015,Zhao2021} and experimental measurements~\cite{Muta2005,Yu2008,Martelli2018};
(c) room-temperature phonon dispersion calculated using SCP and SCHA approaches, where squares denote the SCP frequencies from Ref.~\cite{Tadano2015} for phase-transition soft modes;
(d) temperature-dependent phonon dispersions from 0 to 1000~K using SCP calculations.
}
\label{fig:STO}
\end{figure*}

\subsection*{Application: cubic strontium titanate}
Perovskites are one of the most prominent class of materials with diverse technological applications, such as piezoelectricity, ferroelectricity and superconductivity.
They are also typical strongly anharmonic systems and display multiple structural instabilities driven by lattice anharmonicity.
We further demonstrate the usage of the \texttt{Pheasy} code on the paradigmatic perovskite---SrTiO$_3$, which exhibits a cubic-to-tetragonal phase transition around 110~K~\cite{Cowley1964,Shirane1969}.
To extract IFCs of cubic SrTiO$_3$, the training and testing datasets are prepared directly using the finite-temperature displacements at 300 K generated by the SCHA method.
It also employs Eq.~\eqref{eq:covar_mat} to stochastically populate ensembles, and the effective harmonic IFCs are updated iteratively by performing the minimization of vibrational free energy until a convergence is reached.
More details of the SCHA methodology and implementation will be presented in a forthcoming work, whose underlying physical principles can be found in Ref.~\cite{Monacelli2021}.
After the SCHA convergence is achieved, we use the last two ensembles as the training and testing datasets, respectively, each containing 128 $2\times2\times2$ supercell of the cubic phase.
Then, a sixth-order cluster expansion is performed to represent the potential energy surface of cubic SrTiO$_3$, with a cutoff distance of the fifth-nearest neighbor ($\sim$6.1~\r{A}) for all anharmonic terms beyond third order.
Also, we consider the same many-body interactions as in the case of 2D WS$_2$ for fourth- to sxith-order IFCs.
The resulting numbers of independent IFC parameters from second to sixth order are 45, 698, 2215, 43 and 125, respectively.
Fig.~\ref{fig:STO}(a) demonstrates the accuracy of the adaptive-LASSO extracted IFCs for predicting interatomic forces, and a good agreement with direct DFT calculations is found with a RMSE of 14.442~meV/\r{A}.

When conventional harmonic approximation is used, the calculated phonon spectrum of cubic SrTiO$_3$ is know to exhibit imaginary frequencies due to the dynamical instability, as shown by the dashed lines in Fig.~\ref{fig:STO}(c).
The cubic phase is located at the saddle point of the double-well potential of SrTiO$_3$, which is a local maximum with a negative second-order derivative.
Moving atoms according to the eigen-displacement of soft modes will reduce the total energy and result in the stable tetragonal phase, which is consistent with the observation that the cubic phase is only dynamically stabilized above 110~K with lattice anharmonicity taken into account.
Therefore, we adopt our SCP approach in Eq.~\eqref{eq:covar_mat} to obtain the anharmonicity-renormalized phonon dispersions of cubic SrTiO$_3$ by including the lowest-order loop diagram of phonon self-energy.
Fig.~\ref{fig:STO}(d) shows the temperature-dependent phonon dispersions from 0 to 1000~K, whose frequencies are now all real except for the zero-temperature one.
Within the picture of SCP, the zero-temperature results also contain anharmonic effects because of quantum zero-point energy, which is able to stabilize the two soft modes at the $\Gamma$ and M points but not the antiferrodistortive mode at the R point.
Importantly, we note that the second-order phase transition cannot be observed with the loop self-energy alone, as reflected by a fully stable phonon spectrum of any non-zero temperature.
This holds by definition that the displacement-displacement correlation function [ Eq.~\eqref{eq:covar_mat}] entering the self-consistent equation~\eqref{eq:SCP_loop} must be always positive-definite at a given temperature.
It is the negative-semidefinite bubble diagram of cubic anharmonicity that is responsible for the second-order phase transition~\cite{Bianco2017}.
However, Tadano et al.~\cite{Tadano2015} observed the cubic-to-tetragonal phase transition of SrTiO$_3$ at 220~K by performing SCP calculations, which should be ascribed to the use of an incommensurate phonon grid (i.e. $12\times12\times12$) for evaluating the loop self-energy, while their supercell dimension was only $2\times2\times2$.
Using the same $12\times12\times12$ $\mathbf{q}$-grid, we obtain a critical temperature between 200 to 250 K, in agreement with Tadano et al.~\cite{Tadano2015} (see Fig.~S6 of the Supplementary Information).

On the top of SCP frequencies in Fig.~\ref{fig:STO}(d), we calculate the temperature-dependent LTC of cubic SrTiO$_3$ by considering both 3ph and 4ph scattering.
Fig.~\ref{fig:STO}(b) shows that our calculated LTCs are underestimated across the entire temperature range compared to experimental results~\cite{Muta2005,Yu2008,Martelli2018}, which should be attributed to the use of a relatively small $2\times2\times2$ supercell.
As demonstrated in our previous study~\cite{Han2023}, when employing a larger $4\times4\times4$ supercell, the calculated LTCs from the Boltzmann transport equation were in good agreement with experimental measurements.
Indeed, our 3ph LTC agrees very well with a recent calculation by Zhao et al.~\cite{Zhao2021} from 200 to 1000~K.
When including 4ph processes, our LTC decreases from 9.8 to 8.1 W/mK at 300~K, and a larger drop was observed by Zhao et al.~\cite{Zhao2021}.
Besides, using the Wigner transport formalism~\cite{Simoncelli2019}, we find that the coherence contribution is negligible in cubic SrTiO$_3$ and has therefore been omitted here.

\subsection*{General guidelines for force constant extraction}
Although the use of machine-learning approaches has dramatically simplified IFC calculations, the extraction of IFCs is still not a trivial task, particularly for high-order terms.
Diverse approaches (flavors) for extracting IFCs are recorded in the literature, making the calculated phonon properties highly dependent on the employed extraction scheme.
For instance, the predicted LTC of strongly anharmonic solids can vary significantly among different studies, due to the large uncertainty in the extracted anharmonic IFCs.
Recently, Li et al.~\cite{Li2023} systematically examined the impacts of five different IFC extraction methods on the calculated LTC of Tl$_3$VSe$_4$, revealing a significant variation in the results; however, the optimal scheme for extracting IFCs is not yet identified and remains elusive.
Therefore, it is imperative to determine the most appropriate approach for extracting IFCs to ensure the consistency, accuracy and reliability of the derived phonon-related properties.

Building on our prior studies and rigorous benchmarks on anharmonic lattice dynamics of cubic SrTiO$_3$, we try to identify such optimal approach for IFC extraction and offer five general guidelines in using the \texttt{Pheasy} code:
\begin{enumerate}
    \item if only harmonic IFCs are needed, the use of a small displacement of 0.01~\r{A} is generally a good choice, which has been a common standard adopted in many Phonon codes such as \texttt{Phonopy}~\cite{Togo2015,Togo2023a,Togo2023b} and \texttt{Alamode}~\cite{Tadano2014}.
    The obtained phonon dispersions are in excellent agreement with those from DFPT calculations which are exact in the perturbative limit (see the Supplementary Note 3 for our example on NaCl);

    \item when extracting only third-order IFCs, we recommend to use a small displacement of 0.03~\r{A} and displace all atoms in the supercell along random directions, which should be applicable for most of crystals.
    Remember that there is always a \emph{renormalization effect} from the higher orders into the low ones given the same parity.
    For example, the extracted cubic IFCs are not fully bare if fifth-order terms are not present in the fitting.
    In the same manner, when extracting second- and third-order IFCs at once will yield harmonic ones with the renormalization effects from the fourth order.

    \item a common magnitude of small displacement for extracting fourth-order IFCs does not exist, to our knowledge.
    In this scenario, one can apply a relatively large displacement with a sixth-order expansion to extract bare third- and fourth-order IFCs simultaneously.
    The more preferable alternative is to use finite-temperature displacements generated through a random sampling of quantum canonical ensembles in Eq.~\eqref{eq:covar_mat}, as they represent physical low-energy configurations.
    Once quintic and sextic terms are present, the extracted cubic and quartic IFCs should generally not be sensitive to the specific displacement magnitude, similar to our procedure to extract high-order multipoles (e.g. quadrupoles and octupoles)~\cite{Lin2024}.
    Hence, it is highly recommended to include at least onsite and two-body terms of the fifth and sixth orders when extracting third- and fourth-order IFCs.

    \item when finite-temperature trajectories are used to extract temperature-dependent IFCs,  effective harmonic IFCs should be always fitted first and separately.
    This is to ensure the contribution from the loop diagram of quartic anharmonicity is correctly incorporated into the second-order terms [see Eq.~\eqref{eq:eff_2nd_IFCs}].
    Since the equilibrium structure is not allowed to update during IFC extraction, the effects of the tadpole diagram of cubic anharmonicity are not taken into account.
    Then, temperature-dependent third- and fourth-order IFCs are obtained by fitting to the residual interatomic forces after subtracting the contribution from effective harmonic ones, which can also be set to the SCP solutions.
    Extra attention should be paid when using AIMD trajectories as in the TDEP method, where the bubble diagram's contribution is already included into the effective harmonic IFCs~\cite{Monacelli2025}.

    \item The simultaneous extraction of the IFCs of all orders in the expansion, known as the \emph{one-shot} flavor, is recommended as the most appropriate approach for anharmonic lattice dynamics and thermal transport calculations.
    Particularly, one should avoid using the \emph{cocktail} flavor, which only fits anharmonic IFCs with the harmonic terms fixed at the ones from the DFPT or finite-displacement calculations.
\end{enumerate}
Here, we further rationalize the above point (v) by benchmarking the room-temperature phonon dispersion of cubic SrTiO$_3$ calculated using SCP with that from the SCHA method; the latter serves as the reference results, since SCHA directly computes effective harmonic IFCs without resorting to fourth-order IFCs.
As shown in Fig.~\ref{fig:STO}(c), the SCP results with fourth-order IFCs from the one-shot flavor align perfectly with the SCHA calculations, whereas the SCP spectrum with those from the cocktail flavor deviates.
The bare harmonic IFCs used in the cocktail flavor are obtained from DFPT, where we have exaggerated the difference by adopting an inconsistent setup, i.e. a $2\times2\times2$ $\mathbf{q}$-grid and a $4\times4\times4$ $\mathbf{k}$-grid.
Since our force calculations for extracting anharmonic IFCs are based on a $2\times2\times2$ supercell with a $3\times3\times3$ $\mathbf{k}$-grid (see the ``Methods'' section), the corresponding harmonic terms are in principle equivalent to the ones from the DFPT calculations using a $2\times2\times2$ $\mathbf{q}$-grid and a $6\times6\times6$ $\mathbf{k}$-grid.
This suggests that the cocktail approach is only valid, when one is able to ensure the obtained harmonic IFCs from the DFPT or finite-displacement method are fully equivalent to the actual second-order coefficients of the potential energy, given the same computational setup.
However, such requirement is difficult to fulfill, especially for harmonic IFCs from the finite-displacement method, because the choice of a small displacement of 0.01~\r{A} does not always give the exact derivatives of the potential energy~\cite{Janssen2016}.
Any small deviation in second-order IFCs can result in the magnitude imbalance during the extraction of anharmonic IFCs within the cocktail flavor, and the subsequent SCP and LTC calculations will also be affected.
This fact may potentially explain the much lower LTC of SrTiO$_3$ when 4ph scattering was activated in Ref.~\cite{Zhao2021}, where the cocktial scheme was employed to extract anharmonic IFCs.
Therefore, we recommend to adopt the one-shot approach to extract the IFCs of all orders at once.

\section*{Discussion}
With the demonstrated functionalities and capabilities in this work, \texttt{Pheasy} has been an efficient and robust code for addressing phonon-mediated physics.
Most notably, it significantly accelerates the evaluations of high-order anharmonic IFCs by over two orders of magnitude, compared to conventional finite-displacement methods.
Through the extracted generic IFCs, it further calculates a wide range of phonon properties, including thermodynamic and anharmonic properties (see the Supplementary Note 1 for details), and more features will be available in the near future.
Particularly, it allows to calculate LTCs of materials under different levels of approximations [see Figs.~\ref{fig:silicon}(d) and \ref{fig:ws2}(d)], which is crucial for strongly anharmonic materials with ultralow LTCs.
In order to guarantee the broad accessibility within the phonon community, we have interfaced \texttt{Pheasy} to the two major electronic structure software, \texttt{Quantum ESPRESSO}~\cite{Giannozzi2009,Giannozzi2017} and \texttt{VASP}~\cite{Kresse1996}, via the atomic simulation environment (\texttt{ASE}) package~\cite{HjorthLarsen2017}.
Using the \texttt{ASE} interfaces, it can be further integrated easily with other major electronic structure codes that provide the calculations of total energy, interatomic forces and stress tensor, such as \texttt{Abinit}~\cite{Gonze2016,Gonze2016} and \texttt{Siesta}~\cite{Soler2002}.
Besides, more advanced machine-learning algorithms can be readily incorporated into \texttt{Pheasy} via the \texttt{scikit-learn} package~\cite{Pedregosa2011}, providing a large flexibility in selecting optimal models for extracting IFCs of the potential energy surface.
At the time of writing this paper, \texttt{pheasy} has already been widely used in many studies~\cite{Lin2022,Feng2022,Lin2022,Rodriguez2023a,Rodriguez2023b,Zheng2024,AlFahdi2025,Wei2025,Xiong2024,Han2023,Zheng2023,Ganose2025}, thanks to its simple usage and optimized setups.
For example, it has been integrated into the \texttt{atomate2}~\cite{Ganose2025} framework to power the high-throughput calculations of harmonic phonon properties for inorganic crystals.

\section*{Methods}\label{sec:method}
We employ the \texttt{Quantum ESPRESSO} package~\cite{Giannozzi2009,Giannozzi2017} to perform all DFT and DFPT calculations, using the PBEsol~\cite{Perdew2008} and PBE~\cite{Perdew1996} functionals with the SSSP efficiency pseudopotentials library (v1.3)~\cite{Prandini2018} for bulk and 2D systems, respectively.
All crystal structures are fully relaxed before IFC calculations with the convergence thresholds of pressure, total energy and forces smaller than $10^{-3}$~kbar, $10^{-6}$~Ry and $10^{-5}$~Ry/Bohr, respectively.
LTC is obtained using the iterative solutions of the linearized phonon Boltzmann transport equation as implemented in \texttt{ShengBTE}~\cite{Li2014} with its extension for computing 4ph scattering~\cite{Han2022}.
When 4ph processes are activated, they are evaluated using the relaxation time approximation, with 3ph scattering rates still calculated iteratively.
A particular sampling method with $10^5$ processes as described in Ref.~\cite{Guo2024} is further used to accelerate the estimation of 4ph scattering, and isotope scattering is included in all cases at natural abundance.
Computational details for each material are further described below.

\paragraph*{Silicon.}\label{sec:method_Si}
The plane-wave energy cutoff is set to 60 Ry with a multiple of 8 for the expansion of wavefunction and charge density, respectively.
While an electronic grid of $14\times14\times14$ is used for the calculations with the primitive unit cell, we sample a shifted $2\times2\times2$ grid in the Brillouin zone for any supercell calculation.
The optimized lattice constant of silicon at the PBEsol level is 5.433~\r{A}, and a $30\times30\times30$ $\mathbf{q}$-grid is used to converge the LTC.

\paragraph*{Tungsten disulfide monolayer.}\label{sec:method_WS2}
We choose a plane-wave energy cutoffs as 90 and 900~Ry, respectively, for the expansion of wavefunction and charge density.
All DFT simulations with a single primitive unit cell adopt a $24\times24\times1$ electronic grid, while a shifted $2\times2\times1$ one is used for all supercell calculations.
As a 2D material, we create a vacuum of 20~\r{A} to eliminate spurious interactions in the out-of-plane direction, with the 2D Coulomb cutoff developed by Sohier et al.~\cite{Sohier2017b}.
Our optimized lattice constant of WS$_2$ using the PBE functional is 3.188~\r{A}.
In LTC calculations, phonon scattering is computed on a $90\times90\times1$ $\mathbf{q}$-grid and a monolayer thickness of 6.16~\r{A}~\cite{Schutte1987} is used.

\paragraph*{Cubic strontium titanate.}\label{sec:method_STO}
DFT simulations are performed using a plane-wave energy cutoff of 100~Ry and an $11\times11\times11$ electronic grid, with a multiple of 10 for the expansion of charge density.
We use a $3\times3\times3$ $\mathbf{k}$-grid for all supercell calculations.
Our PBEsol lattice constant of cubic SrTiO$_3$ is 3.890~\r{A}, in good agreement with its experimental value of 3.905~\r{A} at 293~K~\cite{Okazaki1973}.
Phonon properties and LTCs are calculated using a $12\times12\times12$ $\mathbf{q}$-grid, where the analytic correction is activated by the Ewald summation technique~\cite{Gonze1997a} with the Born effective charge and dielectric permittivity tensors from DFPT.

\section*{Data availability}
All data needed to reproduce this study, including optimized crystal structures, DFT inputs, pseudopotentials, interatomic force constants and related output data, will be available on the Materials Cloud Archive.

\section*{Code availability}
The codes \texttt{Quantum ESPRESSO}, \texttt{Phonopy}, \texttt{ShengBTE} and \texttt{FourPhonon} used in this study are all open-source software, available in their respective websites.
The developed \texttt{Pheasy} code is also open-source under GNU General Public License v3.0 and will be made available soon.

\section*{Acknowledgements}
We would like to thank Yanbo Li, Lorenzo Monacelli and Jiongzhi Zheng for useful discussions.
This work is supported by the Sinergia project of the Swiss National Science Foundation (grant number CRSII5\_189924).
The computational time has been provided by the Swiss National Supercomputing Centre (CSCS) under project ID mr33, and the IMX cluster at EPFL.

\section*{Author contributions}
C.L. and B.X. initialized the project.
C.L. wrote the \texttt{Pheasy} package, and J.H. developed the algorithm for null space calculations. 
C.L. performed all calculations, supervised by N.M. and B.X.
All authors analyzed the results and contributed to writing the manuscript.

%\vspace*{0.1cm}
\section*{Competing interests}
The authors declare no competing interests.

\bibliography{Bibliography}

\onecolumngrid

\ifarXiv
    

\section*{Supplementary material}

# Supplementary information: First-principles phonon physics using the `Pheasy` code

Changpeng Lin,[1, *] Jian Han,[2] Ben Xu,[3, †] and Nicola Marzari[1, 4]

[1]*Theory and Simulation of Materials (THEOS), and National Centre for Computational Design and Discovery of Novel Materials (MARVEL), École Polytechnique Fédérale de Lausanne, 1015 Lausanne, Switzerland*
[2]*State Key Laboratory of New Ceramics and Fine Processing, School of Materials Science and Engineering, Tsinghua University, 100084 Beijing, People's Republic of China*
[3]*Graduate School of China Academy of Engineering Physics, 100088 Beijing, People's Republic of China*
[4]*Laboratory for Materials Simulations, Paul Scherrer Institut, 5232 Villigen PSI, Switzerland*
(Dated: August 1, 2025)

**Contents**

---

[*] changpeng.lin@epfl.ch
[†] bxu@gscaep.ac.cn

**Supplementary Note 1: General formulation of lattice dynamics**

In this section, we review and provide the fundamental equations underlying the lattice dynamics of crystals within the harmonic approximation. Then, the expressions used in Pheasy for computing major harmonic and anharmonic properties are also summarized, including thermodynamic functions, thermal mean square displacement, mode participation ratio and Grüneisen parameter.

The equations of motion for atomic vibrations in crystalline solids can be readily derived from Newton's second law, and its secular equation in reciprocal space reads[1,2]:

$$\omega_\nu^2(\mathbf{q}) e_{\nu,\kappa\alpha}(\mathbf{q}) = D_{\kappa\alpha,\kappa'\beta}(\mathbf{q}) e_{\nu,\kappa'\beta}(\mathbf{q}), \tag{1}$$

where $D_{\kappa\alpha,\kappa'\beta}(\mathbf{q})$ is the dynamical matrix at a given wavevector $\mathbf{q}$, $\kappa$ is the atom index within a unit cell, and $\alpha$ is the Cartesian component. If there are $N_\mathrm{a}$ atoms in the unit cell, by diagonalizing $D_{\kappa\alpha,\kappa'\beta}(\mathbf{q})$, one can obtain $3N_\mathrm{a}$ phonon modes labeled by the momentum $\mathbf{q}$ and band index $\nu$, with the frequency $\omega_\nu(\mathbf{q})$ and the corresponding eigenvector $e_{\nu,\kappa\alpha}(\mathbf{q})$. Thus, $D_{\kappa\alpha,\kappa'\beta}(\mathbf{q})$ is the central quantity in a lattice-dynamical problem, which can be constructed via a Fourier transform of real-space interatomic force constants (IFCs) $\Phi_{\kappa\alpha,\kappa'\beta}(\mathbf{R})$ as

$$D_{\kappa\alpha,\kappa'\beta}(\mathbf{q}) = \sum_{\mathbf{R}} \frac{\Phi_{\kappa\alpha,\kappa'\beta}(\mathbf{R})}{\sqrt{m_\kappa m_{\kappa'}}} e^{i\mathbf{q}\cdot\boldsymbol{\tau}_{\kappa\kappa'}(\mathbf{R})}, \tag{2}$$

where $m_\kappa$ is the mass of the $\kappa$th atom in the unit cell, $\boldsymbol{\tau}_{\kappa\kappa'}(\mathbf{R}) \equiv \boldsymbol{\tau}_{\kappa'}(\mathbf{R}) - \boldsymbol{\tau}_\kappa(\mathbf{0})$ is the relative distance between the $\kappa'$th atom of the unit cell $\mathbf{R}$ and the $\kappa$th atom of the unit cell $\mathbf{0}$, $\boldsymbol{\tau}_{\kappa'}(\mathbf{R})$ and $\boldsymbol{\tau}_\kappa(\mathbf{0})$ are their respective positions, and $\mathbf{R}$ is the position vector of a unit cell. Thanks to the translation invariance of crystals, we have set the reference unit cell at $\mathbf{R} = \mathbf{0}$, and it is sufficient to consider the second-order IFCs $\Phi_{\kappa\alpha,\kappa'\beta}(\mathbf{R}) \equiv \Phi_{\kappa\alpha,\kappa'\beta}(\mathbf{0},\mathbf{R})$ between the atoms $\kappa'$ in all unit cells $\{\mathbf{R}\}$ and the atoms $\kappa$ in the reference unit cell. It is noteworthy that the definition of $D_{\kappa\alpha,\kappa'\beta}(\mathbf{q})$ depends on the adopted phase convention, and another common choice is

$$D_{\kappa\alpha,\kappa'\beta}(\mathbf{q}) = \sum_{\mathbf{R}} \frac{\Phi_{\kappa\alpha,\kappa'\beta}(\mathbf{R})}{\sqrt{m_\kappa m_{\kappa'}}} e^{i\mathbf{q}\cdot\mathbf{R}}, \tag{3}$$

which only involves the unit cell position $\mathbf{R}$. The above two definitions are linked by a similarity transformation, and we adopt the latter case in the Pheasy code. It is important to keep a consistent phase convention, especially when the long-range contributions to dynamical matrix are present. Pheasy implements both the Ewald summation technique[3] and the mixed-space approach[4] to handle long-range electrostatic interactions in the long-wavelength limit. In the Ewald summation, we have considered long-range multipolar interactions up to the second order in phonon momentum $\mathbf{q}$, and see Ref.[5] for details; these include the zeroth-order dipole-dipole terms, the first-order dipole-quadrupole terms, and the second-order dipole-octupole, quadrupole-quadrupole and dielectric-dispersion-mediated dipole-dipole terms. By contrast, only the lowest-order dipole-dipole interactions are available in the mixed-space approach.

With phonon frequencies $\omega_\nu(\mathbf{q})$ and eigenvectors $\mathbf{e}_{\nu,\kappa}(\mathbf{q})$ at hand, we can calculate a plethora of harmonic phonon properties. The phonon density of states (DOS) $g(\omega)$ are defined as

$$g(\omega) = \frac{1}{N_\mathbf{q}} \sum_{\mathbf{q}\nu} \delta[\omega - \omega_\nu(\mathbf{q})], \tag{4}$$

where $\delta(\omega)$ is the Dirac delta function at the frequency $\omega$, and $N_\mathbf{q}$ is the number of $\mathbf{q}$-points sampled in the reciprocal space. The atom-projected density of states (PDOS) $g_\kappa(\omega, \hat{\mathbf{n}})$ are further given by

$$g_\kappa(\omega, \hat{\mathbf{n}}) = \frac{1}{N_\mathbf{q}} \sum_{\mathbf{q}\nu} |\hat{\mathbf{n}} \cdot \mathbf{e}_{\nu,\kappa}(\mathbf{q})|^2 \delta[\omega - \omega_\nu(\mathbf{q})], \tag{5}$$

along the direction specified by the unit projection direction vector $\hat{\mathbf{n}}$. For practical calculations, we approximate the Dirac delta function with a Gaussian function as $\delta(\omega) \approx \frac{1}{\sqrt{2\pi}\sigma} \exp(-\omega^2/\sigma^2)$, and $\sigma$ is a smearing parameter in frequency unit. Besides, the precise calculation of $\delta(\omega)$ can be achieved using the tetrahedron integration method that does not rely on any adjustable parameter, and we employ the improved tetrahedron method proposed by Kawamura et al.[6].

The displacement-displacement correlation function defined in Eq. (15) of the main text can be recast into[7]

$$\langle u_{\kappa\alpha}(\mathbf{R}) u_{\kappa'\beta}(\mathbf{R}') \rangle = \frac{\hbar}{2N_\mathbf{q}} \sum_{\mathbf{q}\nu} \frac{2n_\nu(\mathbf{q},T) + 1}{\omega_\nu(\mathbf{q})} \frac{e_{\nu,\kappa\alpha}(\mathbf{q}) e^*_{\nu,\kappa'\beta}(\mathbf{q})}{\sqrt{m_\kappa m_{\kappa'}}} e^{i\mathbf{q}\cdot(\mathbf{R}-\mathbf{R}')}, \tag{6}$$

where $\hbar$ is the reduced Planck constant, $n_\nu(\mathbf{q}, T) = \frac{1}{\exp[\hbar\omega_\nu(\mathbf{q})/(k_\mathrm{B}T)]-1}$ is the Bose–Einstein distribution function at the given temperature $T$, and $k_\mathrm{B}$ is the Boltzmann constant. Then, the thermal mean square displacements (MSDs) for atoms in the reference unit cell $\mathbf{R} = \mathbf{0}$ can be readily obtained as

$$\left\langle \left| u_{\kappa\alpha} \right|^2 \right\rangle = \frac{\hbar}{2N_\mathbf{q}m_\kappa} \sum_{\mathbf{q}\nu} \frac{2n_\nu(\mathbf{q}, T) + 1}{\omega_\nu(\mathbf{q})} |e_{\nu,\kappa\alpha}(\mathbf{q})|^2, \tag{7}$$

which are just the diagonal elements of Eq. (6). Thermal MSDs can be further projected onto an arbitrary direction via the following expression:

$$\left\langle \left| u_{\kappa\alpha}(\hat{\mathbf{n}}) \right|^2 \right\rangle = \frac{\hbar}{2N_\mathbf{q}m_\kappa} \sum_{\mathbf{q}\nu} \frac{2n_\nu(\mathbf{q}, T) + 1}{\omega_\nu(\mathbf{q})} |\hat{\mathbf{n}} \cdot \mathbf{e}_{\nu,\kappa}(\mathbf{q})|^2. \tag{8}$$

The mode participation ratio $P_\nu(\mathbf{q})$ is an useful quantity to measure the localization of phonon modes in the system, originally defined by Bell and Dean[8]:

$$P_\nu(\mathbf{q}) = \frac{1}{N_\mathrm{a} \sum_\kappa |\mathbf{e}_{\nu,\kappa}(\mathbf{q})|^4}. \tag{9}$$

Specifically, $P_\nu(\mathbf{q})$ ranges from $1/N_\mathrm{a}$ to 1, representing the percentage of atoms participating in a given phonon mode. In other words, $P_\nu(\mathbf{q})$ is of $\mathcal{O}(1)$ when all atoms in the unit cell participate in such vibrational mode, and of $\mathcal{O}(1/N_\mathrm{a})$ when only one atom is involved. Additionally, in the literature, there exists another mass-weighted definition as[9]

$$P_\nu(\mathbf{q}) = \left(\sum_\kappa \frac{|\mathbf{e}_{\nu,\kappa}(\mathbf{q})|^2}{m_\kappa}\right) \Big/ \left(N_\mathrm{a} \sum_\kappa \frac{|\mathbf{e}_{\nu,\kappa}(\mathbf{q})|^4}{m_\kappa^2}\right). \tag{10}$$

If removing $m_\kappa$, it is easy to show that the numerator of Eq. (10) becomes unity, given the orthonormal conditions of eigenvectors: $\sum_\kappa \mathbf{e}_{\nu,\kappa}(\mathbf{q}) \cdot \mathbf{e}^*_{\nu',\kappa}(\mathbf{q}) = \delta_{\nu,\nu'}$, and Eq. (10) is hence reduced to Eq. (9). As shown in Fig. 3(d) for NaCl, these two definitions generally yield different results for any compound with multiple elements, and they are only equivalent to each other for monoelemental solids [see Fig. 2 (d) for silicon].

The vibrational thermodynamic properties, including harmonic internal energy $U$, constant-volume heat capacity $C_V$, Helmholtz free energy $F$ and entropy $S$, are calculated via the following expressions[2]:

$$U = \frac{1}{N_\mathbf{q}} \sum_{\mathbf{q}\nu} \hbar\omega_\nu(\mathbf{q}) \left[\frac{1}{2} + \frac{1}{\mathrm{e}^{\hbar\omega_\nu(\mathbf{q})/(k_\mathrm{B}T)} - 1}\right], \tag{11}$$

$$C_V \equiv \left(\frac{\partial U}{\partial T}\right)_V = \frac{k_\mathrm{B}}{N_\mathbf{q}} \sum_{\mathbf{q}\nu} \left[\frac{\hbar\omega_\nu(\mathbf{q})}{2k_\mathrm{B}T}\right]^2 \mathrm{cosech}^2\left[\frac{\hbar\omega_\nu(\mathbf{q})}{2k_\mathrm{B}T}\right], \tag{12}$$

$$F \equiv -k_\mathrm{B}T \ln Z = \frac{k_\mathrm{B}T}{N_\mathbf{q}} \sum_{\mathbf{q}\nu} \left[\ln\left(2\sinh\left[\frac{\hbar\omega_\nu(\mathbf{q})}{2k_\mathrm{B}T}\right]\right)\right], \tag{13}$$

$$S \equiv -\frac{\partial F}{\partial T} = \frac{k_\mathrm{B}}{N_\mathbf{q}} \sum_{\mathbf{q}\nu} \left\{\frac{\hbar\omega_\nu(\mathbf{q})}{2k_\mathrm{B}T} \frac{1}{\mathrm{e}^{\hbar\omega_\nu(\mathbf{q})/(k_\mathrm{B}T)} - 1} - \ln\left[1 - \mathrm{e}^{-\hbar\omega_\nu(\mathbf{q})/(k_\mathrm{B}T)}\right]\right\}, \tag{14}$$

where $Z$ is the partition function.

The mode Grüneisen parameter is defined as[2]

$$\gamma_\nu(\mathbf{q}) \equiv -\frac{\partial \ln \omega_\nu(\mathbf{q})}{\partial \ln \Omega} = -\frac{\Omega}{\omega_\nu(\mathbf{q})} \frac{\partial \omega_\nu(\mathbf{q})}{\partial \Omega}, \tag{15}$$

with the unit cell volume $\Omega$. Using perturbation theory, $\gamma_\nu(\mathbf{q})$ can be obtained easily via cubic IFCs as[10]:

$$\gamma_\nu(\mathbf{q}) = -\frac{1}{6\omega_\nu^2(\mathbf{q})} \sum_{\mathbf{R}\mathbf{R}'} \sum_{\kappa\kappa'\kappa''} \sum_{\alpha\beta\gamma} \frac{e_{\nu,\kappa\alpha}(\mathbf{q})e^*_{\nu,\kappa'\beta}(\mathbf{q})}{\sqrt{m_\kappa m_{\kappa'}}} \Phi_{\kappa\alpha,\kappa'\beta,\kappa''\gamma}(\mathbf{R}, \mathbf{R}')\tau_{\kappa''\gamma}(\mathbf{R}')\mathrm{e}^{i\mathbf{q}\cdot\mathbf{R}}, \tag{16}$$

where $\Phi_{\kappa\alpha,\kappa'\beta,\kappa''\gamma}(\mathbf{R}, \mathbf{R}') \equiv \Phi_{\kappa\alpha,\kappa'\beta,\kappa''\gamma}(\mathbf{0}, \mathbf{R}, \mathbf{R}')$ are the third-order IFCs. With the mode-dependent heat capacity $C_\nu(\mathbf{q})$, the thermodynamic Grüneisen parameter is further calculated as[2]

$$\gamma = \sum_{\mathbf{q}\nu} \frac{\gamma_\nu(\mathbf{q})C_\nu(\mathbf{q})}{C_V}. \tag{17}$$

**Supplementary Note 2: Harmonic and anharmonic phonon properties of silicon**

Here, we perform additional benchmarks on bulk silicon to validate the accuracy of `Pheasy`'s implementation of basic harmonic and anharmonic properties summarized in the Supplementary Note 1. Computational details can be found in the "Method" section of the main text. We calculate harmonic IFCs of silicon using the small-displacement method, where 5 displaced supercells are generated via randomly moving all atoms with a magnitude of 0.01 Å, as compared to the results from `Phonopy`[11–13]. Similarly, we create 20 randomly displaced configurations with a displacement magnitude of 0.03 Å to obtain the third-order IFCs of silicon, which are further compared with those from `thirdorder.py`[14] to calculate lattice thermal conductivity (LTC) using `ShengBTE`[14].

In Fig. 1, we compare our calculated mode Grüneisen parameters, three-phonon scattering rates and LTCs of silicon with the results using cubic IFCs from `thirdorder.py`. An overall good agreement is achieved with just 10 displacements in `Pheasy`, while conventional finite-displacement approach in `thirdorder.py` requires 124 displaced structures, i.e. a speed-up of ∼12 folds. Furthermore, we plot the phonon spectra, density of states and related harmonic and anharmonic properties of silicon in Fig. 2. Our results using `Pheasy` are in excellent agreement with those calculated from `Phonopy`.

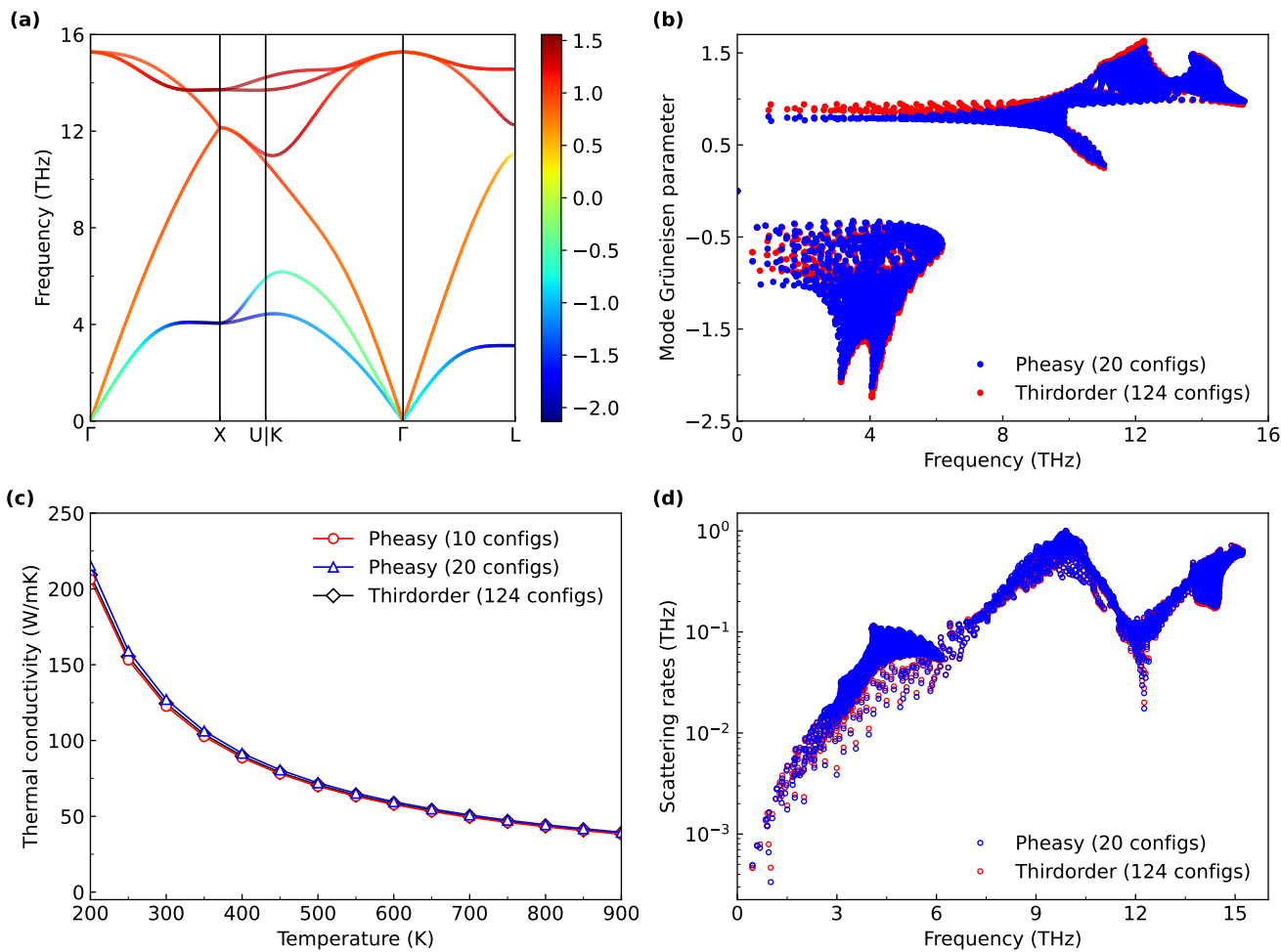

**Supplemental Figure 1**. Grüneisen parameters and thermal transport properties of silicon: (a) phonon dispersions with colors denoting the values of mode Grüneisen parameters; (b) mode Grüneisen parameters as a function of phonon frequencies; (c) temperature-dependent LTCs from 200 to 900 K; (d) three-phonon scattering rates as a function of phonon frequencies. Comparisons are drawn between the results using the third-order IFCs from `Pheasy` and `thirdorder.py`[14].

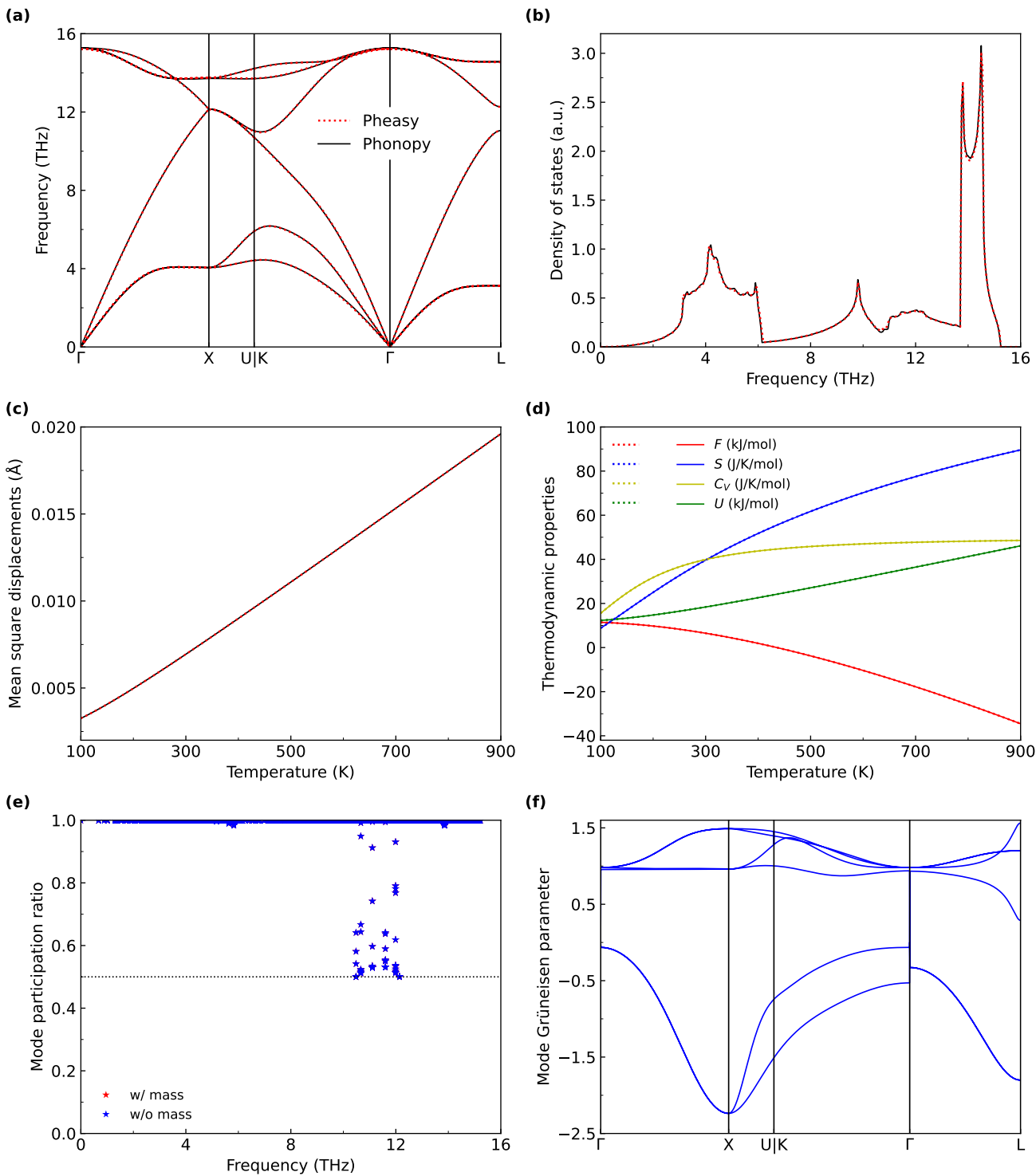

**Supplemental Figure 2**. Harmonic and anharmonic phonon properties of silicon: (a) phonon dispersions; (b) density of states; (c) thermal mean square displacements as a function of temperature; (d) vibrational thermodynamic properties as a function of temperature, including free energy ($F$), entropy ($S$), constant-volume heat capacity ($C_V$) and internal energy ($U$); (e) mode participation ratios as a function of phonon frequencies using two different definitions; (f) mode Grüneisen parameter along the high-symmetry paths. Solid and dotted lines are the results calculated using `Phonopy`[11–13] and `Pheasy`, respectively.

**Supplementary Note 3: Harmonic phonon properties of sodium chloride**

We choose bulk sodium chloride (NaCl) to validate the correctness of our implementation for long-range Coulomb interactions in the `Pheasy` code. DFT calculations are performed using the `Quantum ESPRESSO`[15,16], with the norm-conserving pseudopotentials from PSEUDODOJO[17]. We adopt the Perdew, Burke and Ernzerhof's (PBE) parametrization[18] of the generalized gradient approximation (GGA) for describing the exchange and correlation effects of electrons. The kinetic energy cutoff is set to 160 Ry, and a $12 \times 12 \times 12$ electronic grid is used to sampling the Brillouin zone. Using the same structure optimization criteria as specified in the "Methods" section of the main text, we obtain an optimized lattice constant of NaCl as 5.698 Å. To calculate the real-space IFCs, we use a $3 \times 3 \times 3$ supercell (54 atoms) for the real-space small displacement methods in `Pheasy` and `Phonopy`, which corresponds to a $3 \times 3 \times 3$ **q**-grid in reciprocal-space DFPT calculations. With `Pheasy`, we generate 5 randomly displaced structures with a displacement magnitude of 0.01 Å for all atoms, and we find that 2 configurations are enough to converge harmonic IFCs. Then, DFT calculations with a $2 \times 2 \times 2$ **k**-grid are performed for all supercells to get their interatomic forces, while DFPT calculations are executed with a $6 \times 6 \times 6$ **k**-grid to ensure the consistency of computational setup. The Born effective charge and dielectric tensors of NaCl are also obtained via DFPT, with a denser $12 \times 12 \times 12$ **k**-grid.

The calculated phonon dispersions, atom-projected density of states and related harmonic properties of NaCl are shown in Fig. **3**. Our results from `Pheasy` are in excellent agreement with those from `Quantum ESPRESSO` and `Phonopy`.

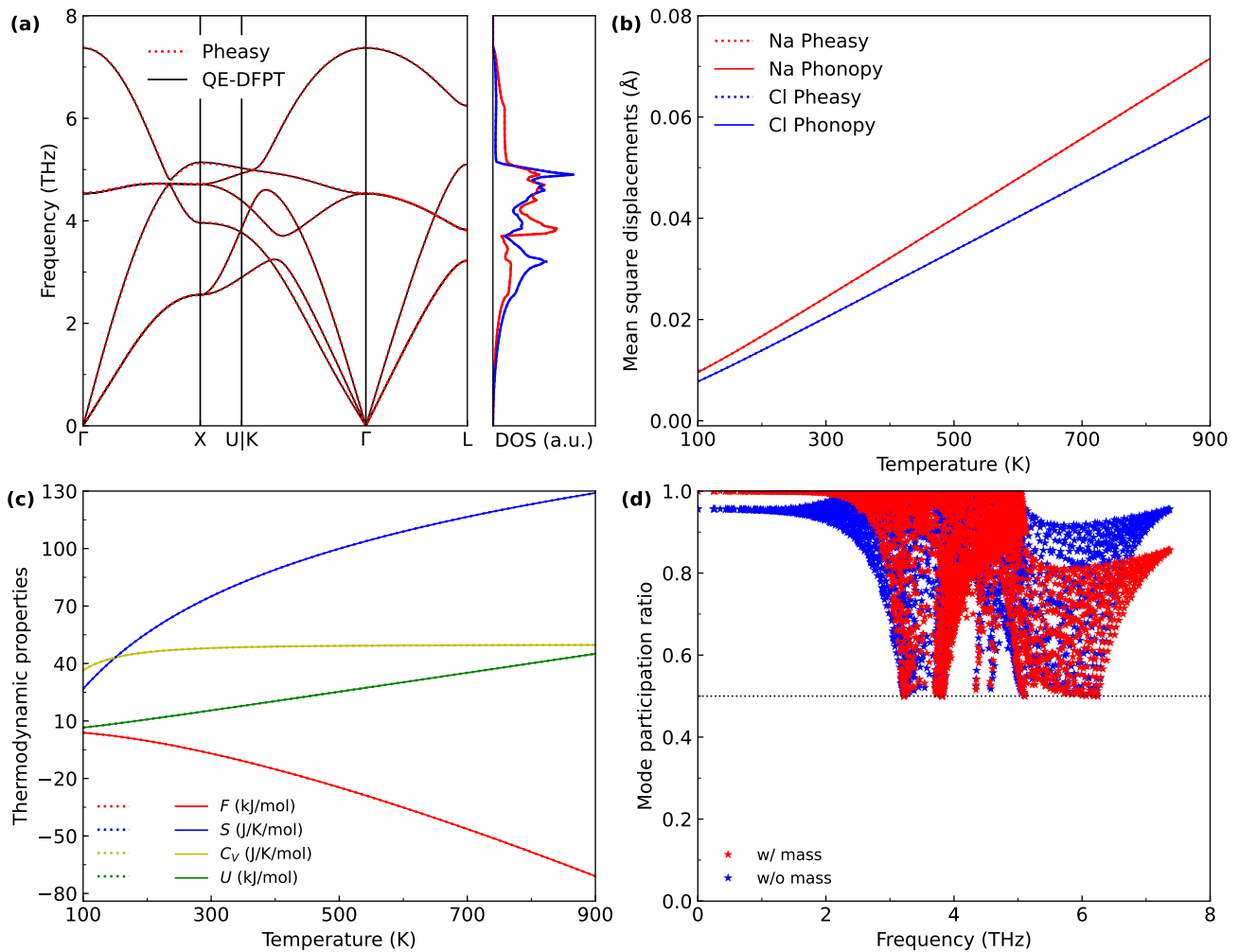

**Supplemental Figure 3**. Harmonic phonon properties of NaCl: (a) phonon dispersions and atom-projected density of states; (b) thermal mean square displacements as a function of temperature; (c) vibrational thermodynamic properties as a function of temperature, including free energy ($F$), entropy ($S$), constant-volume heat capacity ($C_V$) and internal energy ($U$); (d) mode participation ratios as a function of phonon frequencies using two different definitions. Solid and dotted lines are the results calculated using `Phonopy`[11–13] [or `Quantum ESPRESSO` (QE)[15,16]] and `Pheasy`, respectively.

**Supplementary Note 4: Additional figures and tables**

**Supplementary Table 1**. Lattice thermal conductivity of bulk silicon and monolayer WS$_2$ at 300 K calculated with the increasing hierarchy of thermal transport theory, and their self-consistency using different ensembles. We start from the lowest harmonic approximation (HA) to obtain phonon spectra and include only three-phonon (3ph) scattering. The complexity of thermal transport theory increases by taking into account temperature-dependent phonon frequencies (via self-consistent phonons, i.e. SCP) and anharmonic interatomic force constants (IFCs), as well as adding four-phonon (4ph) scattering. All thermal conductivities of silicon and 2D WS$_2$ are calculated based on a $30 \times 30 \times 30$ and a $90 \times 90 \times 1$ **q**-mesh, respectively. The iterative solution of phonon Boltzmann transport equation is achieved only for 3ph scattering, while 4ph scattering is obtained within the relaxation time approximation (RTA). A sampling method[19] using $10^5$ processes is further used to accelerate the estimation of 4ph scattering within RTA, and the thickness of 2D WS$_2$ is set to 6.16 Å[20]. The force-displacement dataset was generated using the quantum canonical sampling of normal modes at 300 K, and there are 64 configurations in each ensemble.

| Materials | Ensemble No. | HA+3ph | HA+3,4ph | SCP+3ph | SCP+3,4ph | SCP+3ph+IFCs@$T$ | SCP+3,4ph+IFCs@$T$ |
|---|---|---|---|---|---|---|---|
| Silicon | 1 | 124.85 | 115.13 | 122.49 | 112.48 | 125.90 | 115.32 |
|  | 2 | 125.40 | 115.54 | 122.33 | 112.18 | 124.57 | 114.18 |
| WS$_2$ | 1 | 243.69 | 206.43 | 241.90 | 205.87 | 238.39 | 201.05 |
|  | 2 | 241.61 | 202.48 | 239.68 | 201.71 | 239.10 | 202.70 |

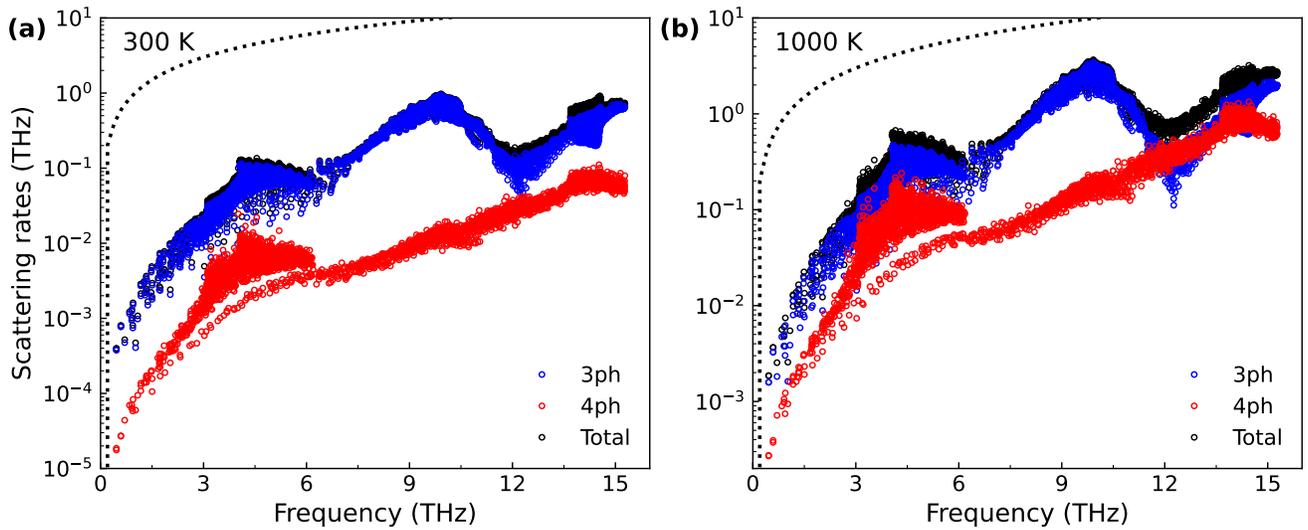

**Supplemental Figure 4**. Three- and four-phonon scattering rates as a function of phonon frequencies in silicon at (a) 300 and (b) 1000 K. The dotted line indicates where phonon frequency equals scattering rate, i.e. the breakdown of quasiparticle approximation.

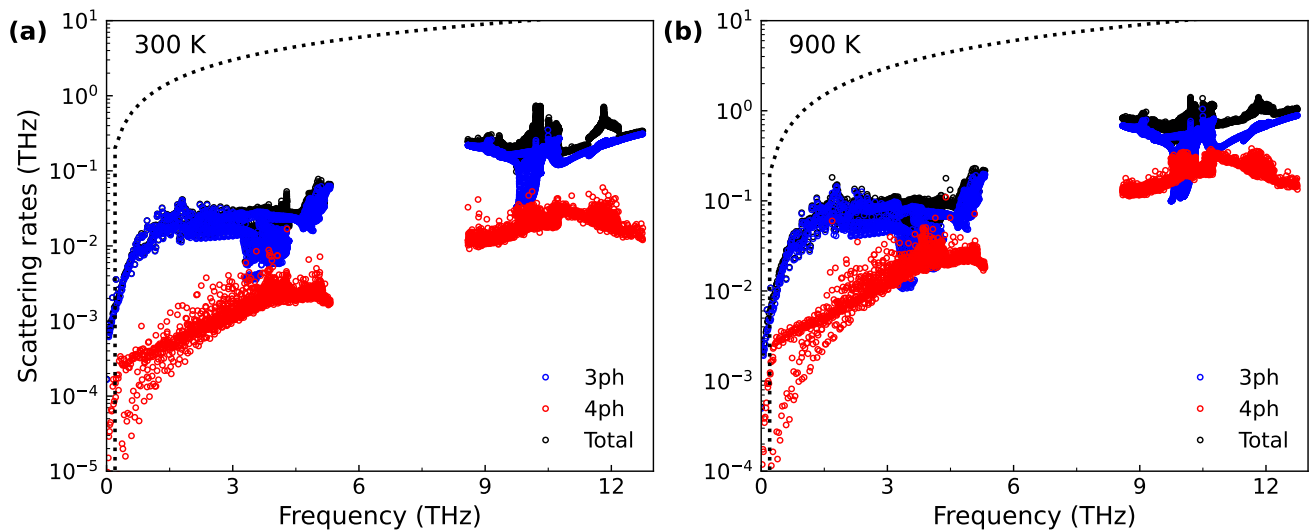

**Supplemental Figure 5**. Three- and four-phonon scattering rates as a function of phonon frequencies in $WS_2$ monolayer at (a) 300 and (b) 900 K. The dotted line indicates where phonon frequency equals scattering rate, i.e. the breakdown of quasiparticle approximation.

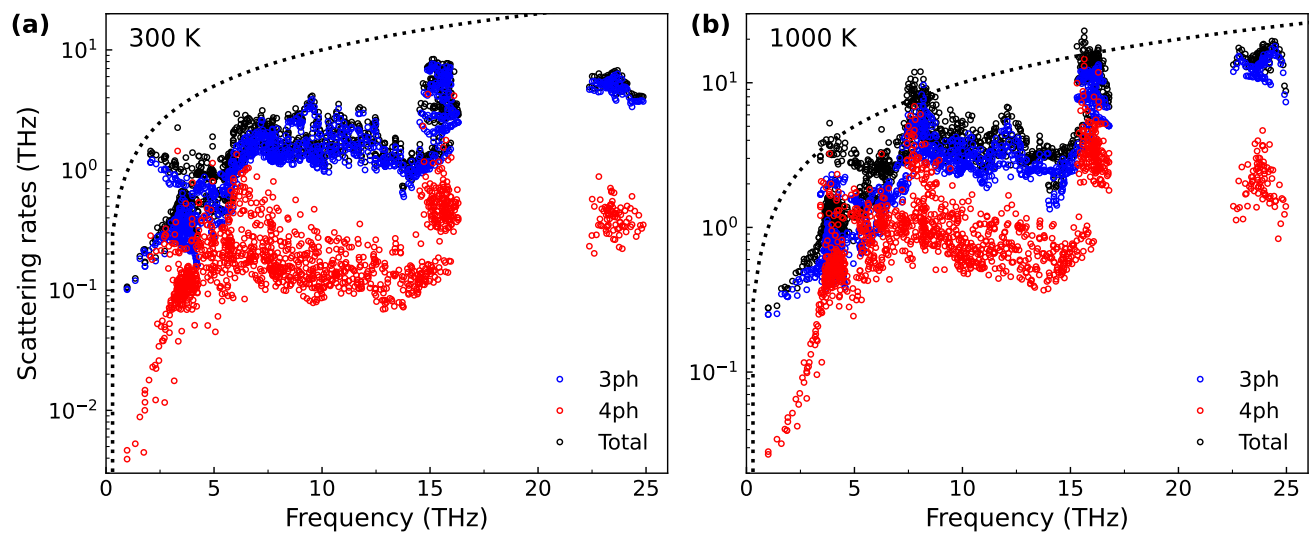

**Supplemental Figure 6**. Three- and four-phonon scattering rates as a function of phonon frequencies in cubic $SrTiO_3$ at (a) 300 and (b) 1000 K. The dotted line indicates where phonon frequency equals scattering rate, i.e. the breakdown of quasiparticle approximation.

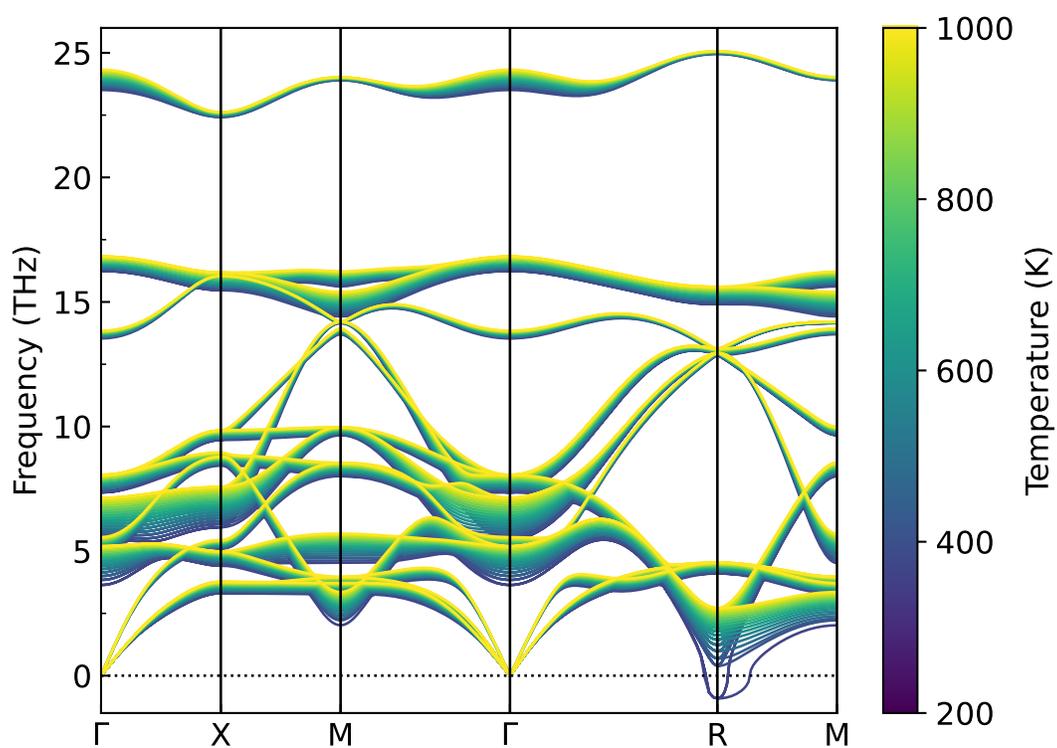

**Supplemental Figure 7**. Temperature-dependent phonon dispersions of cubic SrTiO$_3$ from 200 to 1000 K, calculated using the SCP approach with the loop diagram of phonon self-energy. The SCP equation is solved on a $12 \times 12 \times 12$ **q**-grid, which suggests a phase-transition temperature between 200 and 250K, in agreement with the SCP calculations by Tadano et al.[21].


\fi

\twocolumngrid

\end{document}